\documentclass{nature}

\spacing{1}
\usepackage{booktabs}
\usepackage{xspace}
\usepackage{amsmath,amssymb}
\usepackage{xcolor}
\usepackage{hyperref}
\usepackage{graphicx}
\usepackage[numbers]{natbib}
\usepackage{multicol}
\usepackage{caption}

\graphicspath{{figures/}}

\makeatletter
\let\saved@includegraphics\includegraphics
\AtBeginDocument{\let\includegraphics\saved@includegraphics}
\renewenvironment*{figure}{\@float{figure}}{\end@float}
\makeatother

\newcommand{\hGpc}{\;h^{-1}~{\rm Gpc}}

\newcommand{\hMpc}{{\ifmmode{\;h^{-1}{\rm Mpc}}\else{$h^{-1}$Mpc}\fi}}
\newcommand{\hkpc}{{\ifmmode{\;h^{-1}{\rm kpc}}\else{$h^{-1}$kpc}\fi}}
\newcommand{\hMsun}{{\ifmmode{\;h^{-1}{\rm {M_{\odot}}}}\else{$h^{-1}{\rm{M_{\odot}}}$}\fi}}
\newcommand{\Msun}{\;\rm {M_{\odot}}}
\newcommand{\Mstar}{{\ifmmode{\;M_{*}}\else{$M_{*}$}\fi}}
\newcommand{\Mhalo}{{\ifmmode{\,M_{\rm halo}}\else{$M_{\rm halo}$}\fi}}
\newcommand{\ltsima}{$\; \buildrel < \over \sim \;$}
\newcommand{\gtsima}{$\; \buildrel > \over \sim \;$}
\newcommand{\lsim}{\lower.5ex\hbox{\ltsima}}
\newcommand{\gsim}{\lower.5ex\hbox{\gtsima}}

\newcommand{\theth}{{\sc The Three Hundred}}

\newcommand{\simba}{\textsc{Gizmo-SIMBA}}
\newcommand{\gadgetx}{\textsc{Gadget-X}}
\newcommand{\gadgetmusic}{\textsc{Gadget-MUSIC}}

\newcommand{\MCNN}{M_{\text{CNN}}}
\newcommand{\Mtrue}{M_{\text{true}}}
\newcommand{\MPlanck}{M_{\text{SZ}}^{\text{Planck}}}

\newcommand{\bP}{b_{\text{P}}}
\newcommand{\cleandata}{\emph{Clean mock data set}}
\newcommand{\mockdata}{\emph{Planck mock data set}}
\newcommand{\realdata}{\emph{Planck real data set}}
\newcommand{\goldendata}{\emph{Golden sample}}

\newcommand{\MSZ}{M_{\text{SZ}}}

\newenvironment{Figure}
  {\par\medskip\noindent\minipage{\linewidth}}
  {\endminipage\par\medskip}

\title{A Deep Learning Approach to Infer Galaxy Cluster Masses from  Planck Compton$-y$ parameter maps}
\author{
Daniel de Andres,$^{1}$\thanks{E-mail: daniel.deandres@uam.es} ~ 
Weiguang Cui,$^{1,2}$\thanks{E-mail: weiguang.cui@uam.es} Florian Ruppin,$^3$ Marco De Petris,$^4$ Gustavo Yepes,$^1$ Giulia Gianfagna,$^{4,5}$ Ichraf Lahouli,$^6$ Gianmarco Aversano,$^6$ Romain Dupuis,$^6$ Mahmoud Jarraya,$^6$
and Jesús Vega-Ferrero$^{7}$
}

\begin{document}
\maketitle

\begin{affiliations}
\item Departamento de Física Teórica and CIAFF, Modulo 8 Universidad Autónoma de Madrid, 28049 Madrid, Spain.
\item Institute for Astronomy, University of Edinburgh, Blackford Hill, Edinburgh, EH9 3HJ, UK
\item Kavli Institute for Astrophysics and Space Research, Massachusetts Institute of Technology, Cambridge, MA 02139, USA
\item Dipartimento di Fisica, Sapienza Universitá di Roma, Piazzale Aldo Moro, 5-00185 Roma, Italy 
\item INAF - Istituto di Astrofisica e Planetologia Spaziali, via Fosso del Cavaliere 100, I-00133 Roma, Italy
\item EURANOVA, Mont-Saint-Guibert, Belgium
\item Instituto de Astrofísica de Canarias (IAC) La Laguna, 38205, Spain
\end{affiliations}

\begin{abstract}

Galaxy clusters are useful laboratories to investigate the evolution of the Universe, and accurately measuring their total masses allows us to constrain important cosmological parameters. However, estimating mass from observations that use different methods and spectral bands introduces various systematic errors. This paper evaluates the use of a Convolutional Neural Network (CNN) to reliably and accurately infer the masses of galaxy clusters from the Compton-y parameter maps provided by the Planck satellite. The CNN is trained with mock images generated from hydrodynamic simulations of galaxy clusters, with Planck’s observational limitations taken into account. We observe that the CNN approach is not subject to the usual observational assumptions, and so is not affected by the same biases. By applying the trained CNNs to the real Planck maps, we find cluster masses compatible with Planck measurements within a 15\% bias. Finally, we show that this mass bias can be explained by the well known hydrostatic equilibrium assumption in Planck masses, and the different parameters in the Y500-M500 scaling laws. This work highlights that CNNs, supported by hydrodynamic simulations, are a promising and independent tool for estimating cluster masses with high accuracy, which can be extended to other surveys as well as to observations in other bands.
\end{abstract}


\begin{multicols}{2}
\section{Introduction}
Galaxy clusters are the biggest gravitational bound objects in the Universe. Dark matter is the main component in galaxy clusters, which amounts to around 80\% of their total mass and therefore, it is responsible for the gravitational collapse of structures. Structures can then grow hierarchically, merging with other halos to form massive clusters within the range of $10^{14}$-$10^{15}$\(M_\odot\) \citep[for a full review see e.g.][]{kravtsov2012formation}. Moreover, about 8\% of the clusters' mass  corresponds to galaxies and the remaining 12\% is diffused as hot gas between galaxies, i.e. the intra-cluster medium (ICM). An accurate estimation of galaxy cluster masses is of paramount importance in cosmology due to the fact that one can constrain different cosmological parameters through the halo mass function \citep[e.g.][]{aghanim2020planck}. 

However, the total mass of a cluster is not a direct observable in the images from telescopes. It can be only inferred by different approaches: for example, the dynamics of the member galaxies \citep{Biviano2006}; ICM radial profiles from X-ray or Sunyaev-Zel’dovich (SZ) observations with the  assumption of  hydrostatic equilibrium (HE) \citep{kravtsov2012formation}; weak gravitational lensing (WL) analysis \citep{Becker2011}.
Alternatively, suitable observational proxies can be selected among clusters physical quantities strictly related to the mass of the object under the self similarity assumption \citep{bryan1998}. Nevertheless, in all the listed methods to infer the mass we have to face with the problem of mass bias -- the derived mass is systematically different from the real cluster mass due to the assumed approximations in each approach.
The presence of the mass bias has an impact on the inference of cosmological parameters and in particular, the cosmic matter density $\Omega_m$ and the normalisation of the matter power spectrum $\sigma_8$. Currently, the value of the bias, $b = \Delta M/M$ where $\Delta M$ is the mass difference between the estimated mass and the real mass, needed to reconcile CMB constraints with thermal SZ (tSZ) cluster counts is $(1-b)=0.58\pm 0.04$ \citep{planck2016b}. Such a large value for the bias is not consistent with almost any  of the estimates based on X-ray, SZ and WL observations, on average,  all are around $(1-b)=0.80\pm 0.08$ \citep{salvati2018}. Large-scale hydrodynamic simulations play an important role regarding the determination and calibration of the mass bias. However, even several data-sets with different physical processes included, particle resolutions and number of objects result inconsistent with the Planck bias requirement, (see e.g. \cite{giuliabias} for a recent review).

Machine Learning (ML) \citep[e.g.][]{bishop2006pattern} algorithms allow us to analyse data and make predictions without assuming any previous known behaviour, i.e. data driven science. The rapid growth in data complexity in astronomy encourages the development of these techniques, where a wide variety of ML models have been studied so far \citep[e.g.][]{baron2019machine}. Deep Learning \citep[e.g.][]{lecun2015deep,Goodfellow-et-al-2016} is a ML tool that makes use of multilayer perceptrons (MLPs), also known as feedforward neural networks, with numerous ``deep" hidden layers. In particular, Convolutional Neural Networks \citep[CNNs;][]{lecun1989generalization} are a type of neural network that use convolutions for processing data that show a known grid-like topology.
Moreover, recent studies have shown that Deep Learning methods can be used for inferring galaxy cluster masses directly from mock X-ray images \citep[][]{ntampaka2019deep}; mock SZ images \citep[][]{gupta2020massSZ}; CMB cluster lensing \citep[][]{gupta2020massWL}; a combination of X-ray, SZ and optical mock images \citep[][]{yan2020galaxy}; and from galaxy members dynamics \citep{ho2019robust,kodi2021simulation,ho2021approximate}. These techniques do not rely on any assumption  on the dynamics or the spherical symmetry of the ICM, but rather on the quality of data set. These theoretical studies further suggested a bias free estimation of the cluster mass. 

In this study, we make another step by applying these theoretical works to predict the masses of real galaxy clusters observed through the SZ effect. Particularly, we apply the trained CNNs to the publicly available Second Planck catalogue of Sunyaev-Zel'dovich sources, i.e. PSZ2 catalogue \citep[][]{planckcatalog}, to derive the cluster masses. In order to do that, we analyze a sample of 6765 clusters from the \theth{} \citep[The300;][]{Cui2018} simulation with the same redshift and mass ranges as PSZ2 clusters. Particularly, we train our CNNs using simulated tSZ images aiming at predicting the masses from real Compton-$y$ parameter Planck images. We also compare our results with masses estimated by Planck and determine that masses inferred with our CNN are overall in agreement but showing some discrepancies that might be attributed to the general assumptions used in Planck, such us the hydrostatic equilibrium and the $Y$-$M$ scaling relation. 

To this end, the mock SZ maps have the same noise and beam convolution as the corresponding Planck observations. A summary of the characteristics of each data set used in this work is presented in \autoref{table:mapsummary}. The interested reader can find further information and technical details regarding the generation of these mock observations in the Methods section (\S \ref{section:dataset}). For information concerning the CNN model, training and validation procedure, the choice of redshift bins and error estimations, we refer the reader to the supplement material.

\begin{table*}
\centering
\caption{{\bf Title: Detailed data sets used in this study.} Data properties of the simulated data sets \cleandata{} and \mockdata{} and the observations \realdata{} and \goldendata{} }
\label{table:mapsummary}
\begin{tabular}{ccccc} 
\toprule
Data set             & mock/real & \begin{tabular}[c]{@{}c@{}}Beam smoothing \\ (FWHM 10 arcmin )\end{tabular} & \begin{tabular}[c]{@{}c@{}}Instrumental \\Noise\end{tabular} & \begin{tabular}[c]{@{}c@{}}point source\\ contaminants\end{tabular}  \\ 
\toprule
Original mock data set & mock & no ($5''$) & No & No \\
Clean mock data set  & mock  & yes  & No   & No    \\
Planck mock data set & mock  & yes  & Yes  & No    \\
Planck real data set & real  & yes  & Yes  & Yes   \\
Golden sample        & real  & yes  & Yes  & No    \\
\bottomrule
\end{tabular}
\end{table*}

\section{Results}\label{section:results}

\subsection{Verifying the CNN models}

\begin{figure*}
\centering
\includegraphics[width=0.7\textwidth]{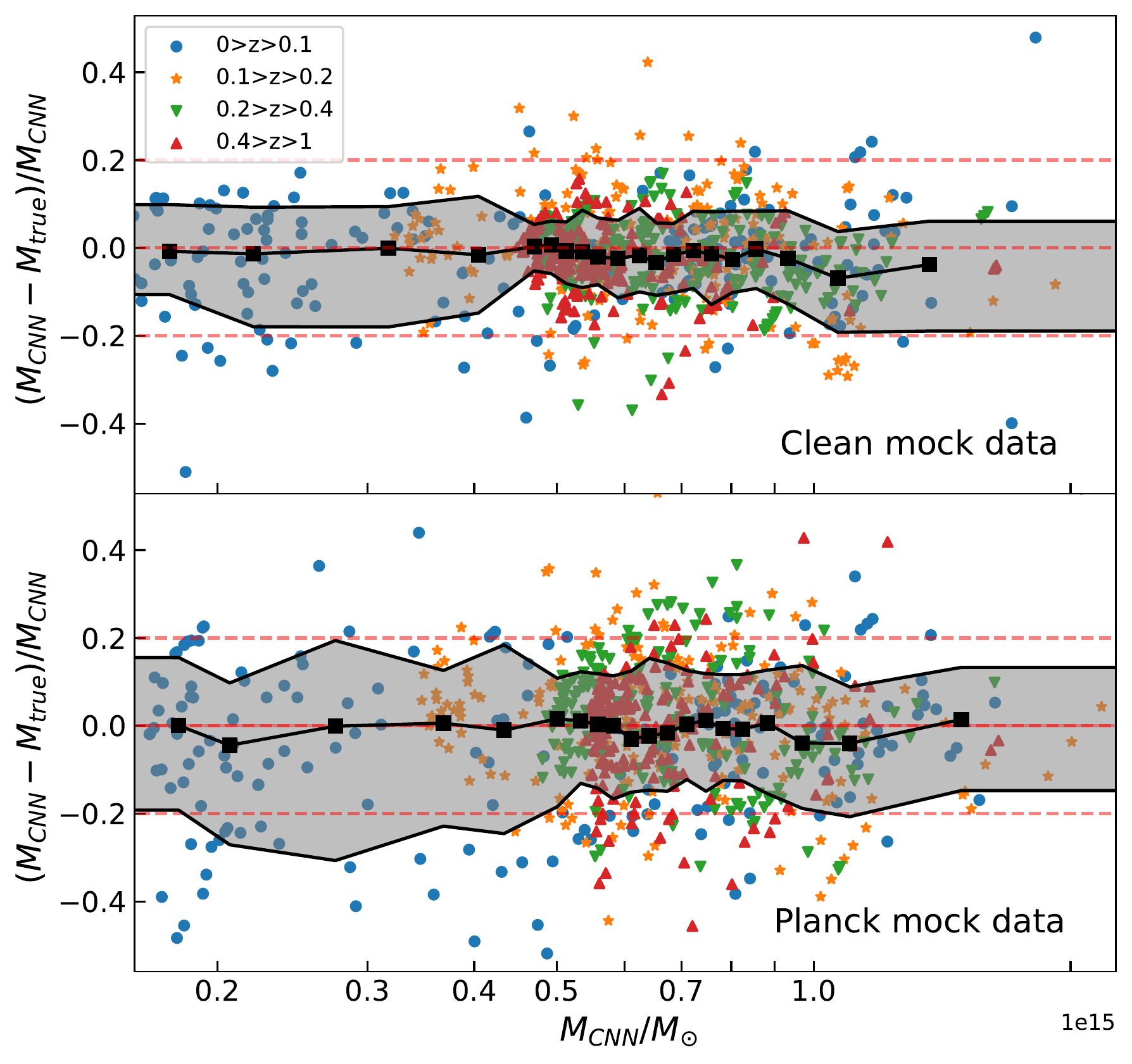}
\caption{{\bf verifying CNN with mock maps.} The relative error $(\MCNN-\Mtrue)/\MCNN$ for \cleandata{} (top) and the \mockdata{} (bottom) as a function of the predicted mass $\MCNN$. Black squares (bin centre) represent median values, while the shaded region is the $16^{\text{th}}-84^{\text{th}}$ percentiles. Furthermore, red dashed lines correspond to the perfect prediction (0 error) and $\pm20\%$ error and different colour points depict different redshift ranges as shown in the legend. A random sample of 200 points per redshift bin is shown but the statistics (median and $16^{\text{th}}-84^{\text{th}}$ percentiles) are computed using the whole test set. The data is binned along $\MCNN$ such that every bin has n=962 $y$-maps.}
\label{fig:preds_simulation}
\end{figure*}

Our estimated CNN masses, $\MCNN$, in comparison with the real cluster mass $\Mtrue$ (quantified with $M_{500}$, i.e. the spherical overdensity halo mass at $500\times \rho_{crit}$, here $\rho_{crit}$ is the critical density of the Universe) are shown in  \autoref{fig:preds_simulation} for simulated clusters using both \cleandata{} and \mockdata{}. \autoref{fig:preds_simulation} shows the relative error as a function of the predicted CNN mass $M_{\text{CNN}}$:
\begin{equation}\label{eq:error}
    \text{err}(M_{\text{CNN}},\Mtrue)=\frac{M_{\text{CNN}}-\Mtrue}{M_{\text{CNN}}}=b_{\text{t}}\text{ ,}
\end{equation}
where $b_{\text{t}}$ can be thought as the bias of the true mass with respect to the $\MCNN$ mass.
As shown in the top panel of \autoref{fig:preds_simulation}, $b_{\text{t}}$ has median values at around -0.02, at the 4 redshift bins and within the whole mass range. The scatter in this CNN estimated mass is within 20\%. Even when training with and applying to the \mockdata{}, it is clear (from the bottom panel of \autoref{fig:preds_simulation}) that the CNN mass is only slightly biased towards a negative value ($\lesssim 5\%$, see the supplementary section D for more discussions). However, the shaded region increases from $-0.02^{+0.09}_{-0.10}$ (standard error of $\pm 0.001$) to $-0.03^{+0.14}_{-0.17}$ (standard error of $\pm 0.002$), which indicates the impact of the instrumental Planck noise in the CNN predictions. We note here that the scatter is comparable to the results from \cite{yan2020galaxy} (see their Fig. 7). Furthermore, we also trained our model to estimate the cluster mass $M_{200}$, the results were poor, in the sense of relative larger scatter in the bias $b_t$, due to the fact that the signal becomes weak at $R>R_{500}$ for the \mockdata{}. While, $M_{200}$ can be estimated using the \cleandata{} with a similar accuracy.

\begin{figure*}
\includegraphics[width=1.\textwidth]{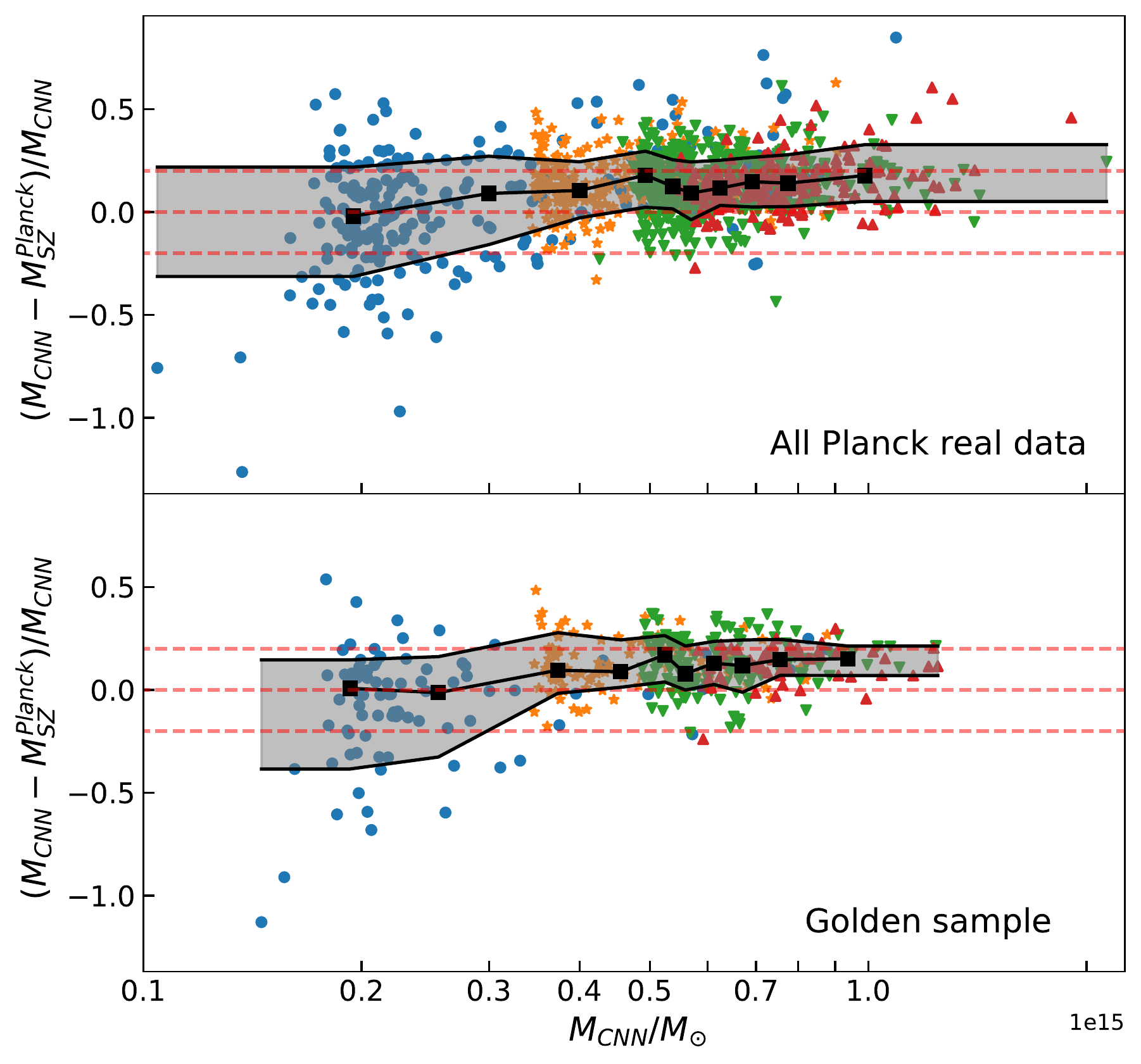}
\centering
\caption{{\bf  Comparing CNN predicted cluster mass with the mass estimated by Planck.} Similar to \autoref{fig:preds_simulation} but for all \realdata{} (top panel) and for the \goldendata{} (bottom panel). The definition of these considered data sets can be found in section §\ref{subsec:mockobservations}. The CNN mass is binned such that every mass bin consists of n=109 clusters in the case of \realdata{} and n=39 clusters in the case of \goldendata{}.}
\label{fig:preds_planck}
\end{figure*}

\subsection{Predicting the Planck cluster masses}

We simply apply the CNNs trained with \mockdata{} to the \realdata{} for predicting their masses (see section §\ref{subsec:mockobservations} for a detailed description of the data sets). The results are shown in \autoref{fig:preds_planck} by presenting the relative errors between our CNN masses and the cluster masses estimated by Planck, $\MPlanck$, as a function of the predicted mass, $\MCNN$. Similarly to \autoref{eq:error}, we define 
\begin{equation}
    \text{err}(M_{\text{CNN}},\MPlanck)= \frac{M_{\text{CNN}}-\MPlanck}{M_{\text{CNN}}} = b_{\text{P}} \text{,}
\end{equation}
here $\bP$ is the bias of Planck masses with respect to the $\MCNN$. Note that the cluster masses estimation from Planck is based on the HE assumption. \cite{Planck2013} has predicted an average mass bias of $1-b = 0.8$. Different to the results in the case of the \mockdata{}, the median value of $\bP$ is clearly biased towards a positive value, $\bP = 0.11^{+0.14}_{-0.15}$ ($\pm 0.005$ of standard error), for massive clusters ($\MCNN/\Msun \gtrsim 4\times 10^{14}$). At lower cluster mass, $\bP \approx -0.03^{+0.24}_{-0.27}$ ($\pm 0.02$ of standard error) which means a consistent cluster mass between our CNN and the Planck estimations. We would like to note that the scatter shown by the shaded region in \autoref{fig:preds_planck} is also inline with the results in the lower panel of \autoref{fig:preds_simulation}. It is clear that there is about 0.1 difference between this bias and the Planck estimated bias. 
 

In the full $\realdata{}$, roughly 2/3 of the clusters have contamination by point-like sources near their centre, which is not present in the simulated maps. To verify whether this is the cause of the mass bias difference between our CNN method and the Planck result, we select a sub-sample of the $\realdata{}$ which does not have any relevant radio source or other contaminants in the cluster centre or the vicinity (within 10~arcmin). Furthermore, radio emission contamination outside of the main halo is substituted with a signal intensity which is compatible with instrumental noise. This sub-sample is named as the \goldendata{} and its result is shown in the bottom panel of figure~\ref{fig:preds_planck}. Although this \goldendata{} contains a smaller number of objects, its median $\bP$ is in a good agreement with the result from the full $\realdata{}$. For the exact values of the biases of the \goldendata{} and the full \realdata{}, we also refer to the supplementary section D (see Supplementary Table 2). Clearly, this bias is not caused by the detected point-like contaminants. Therefore, we investigate other possibilities in the following section in order to explain the difference between $\MCNN$ and $\MPlanck$.


\begin{Figure}
\includegraphics[width=1.1\columnwidth]{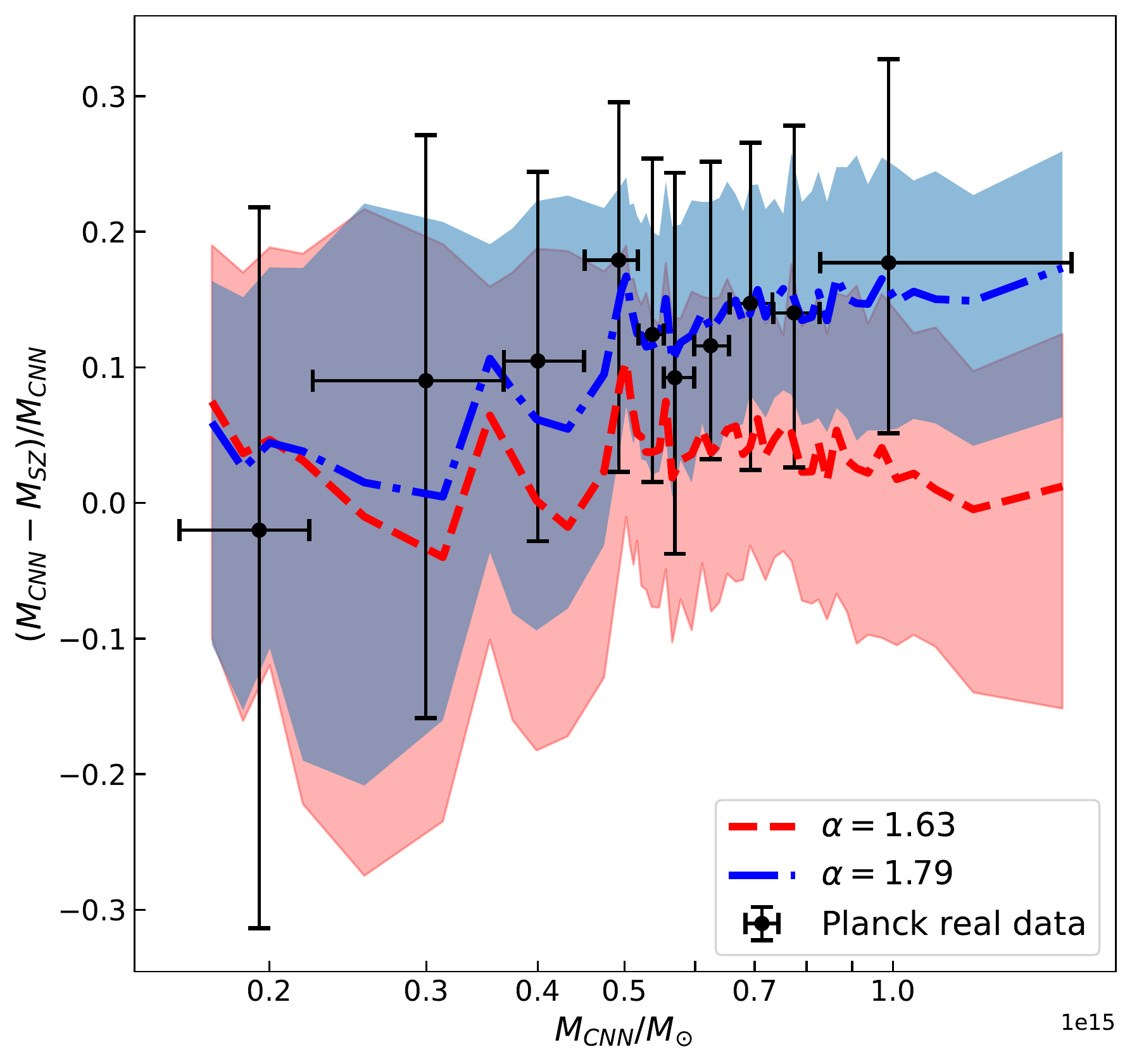}

 \captionof{figure}{{\bf Verifying the bias causes with $Y-M$ relation.} The relative error $(\MCNN-M_{\text{SZ}})/\MCNN$ as a function of the predicted mass $\MCNN$ for different values of the parameter $\alpha$ in \autoref{eq:scaling_relations}. Values from \theth\ self-scaling relation, where $\alpha=1.63$, are shown in red dashed line and the parameter used by Planck $\alpha=1.79$ in blue dot-dashed line. The shaded red and blue regions correspond to the $16^{th}-84^{th}$ percentiles respectively.  Black points represent the median values of $(\MCNN-\MPlanck)/\MCNN$ using Planck real data with an error bar equal to the $16^{th}-84^{th}$ percentile and x-axis errors show the bin range. The CNN mass is binned such that every mass bin consists of n=109 clusters in the case of \realdata{} and n=142 $y\text{-maps}$ in the case of simulated data.}
\label{fig:scalingrelations}
\end{Figure}

\subsection{Understanding the mass bias}

\label{sec-bias}
Limited to our knowledge on the detailed processes of estimating the $\MPlanck$, we perform a simple inference of the cluster mass with the mock $y$ maps to compare with its CNN mass. It is well known that the relation between the integrated Compton-$y$ parameter $Y$ which is proportional to the thermal energy in the ICM \citep{sunyaev1972}, over an aperture of radius $R$, and the mass inside the same aperture, $M$, is a power law. Accordingly, $Y$ is defined as an integral over an aperture subtended by a solid angle $\Omega$ :
\begin{equation}
    Y = \int_{\Omega} y d\Omega \simeq \sum_{i}^{i\in R} y_{i}\Omega_{i}\text{,}
\end{equation}
where $\Omega_{i}$ is the area of the pixel $i$ and the sum is performed over the image pixels inside $R$. Here we focus on quantities integrated inside $R_{500}$ in order to compare our estimation with other masses, e.g. $\MPlanck$. This $Y$-$M$ scaling law can be written as
\begin{multline}\label{eq:scaling_relations}
E(z)^{-2/3}\left[\frac{D_{A}^{2}(z)Y}{10^{-4} \text{Mpc} ^{2}}\right] = B\left[\frac{h}{70}\right]^{-2+\alpha}\\\times\left[ \frac{\MSZ}{6\times 10^{14}\Msun}\right]^{\alpha}\text{,}
\end{multline}
where $D_{A}(z)$ is the angular diameter distance at redshift $z$ and $E(z)= H(z)/H_{0}$ is the redshift evolution of the Hubble $H(z)$ parameter where $h$ is its dimensionless value, i.e. $H= 100$ $h$ ${\rm km/s/Mpc}$. Particularly, the fitted parameters: slope $\alpha$ and normalisation $B$, for \cite{ade2014planck} are $\alpha=1.79\pm0.08$ and $\log (B)=-0.19\pm 0.02$ at $R_{500}$. The estimation of these parameters is based on the cluster masses from a mass-proxy relation from \cite{Kravtsov_2006}.  The normalisation $B$ parameter is similar between \theth\ clusters and the Planck result. However, the slope of this relation in \theth\ is $\alpha=1.63\pm 0.29$, which is compatible with a self similar relation with $\alpha=5/3$ \citep{Cui2018}. Note that the large error in the slope from \theth\ is due to a mass-complete fitting process, interested readers are referred to \citep{Cui2018} for details.

In order to examine whether the difference in the slope of the $Y$-$M$ scaling relation is the cause of the bias, we derive the $Y_{500}$ from the original mock data set. Note that the $R_{500}$ estimated in the AHF catalogue \citep{Knollmann2009} is used here. $\MSZ$ is then converted from $Y$ using \autoref{eq:scaling_relations} with two slopes: $\alpha=1.63$ (\theth) and $\alpha=1.79$ (Planck) based on \autoref{eq:scaling_relations}. 
In addition, to meet Planck results, we applied the same correction factor 1.2 (from $Y_{\text{sph}}$ to $Y_{\text{cyl}}$ \cite{Planck2013}) to the $Y$ from the original mock maps (blue line), while for the red line, we simply adopt the fitting parameter from \cite{Cui2018} which used $Y_{cyl}$. Here, $Y_{\text{cyl}}$ and $Y_{\text{sph}}$ are the integrated Compton-y parameter over a cylindrical and spherical region respectively.
In \autoref{fig:scalingrelations}, we show the relative errors between $\MCNN$ and $\MSZ$ as a function of $\MCNN$ for $\MSZ$ masses estimated through the two different scaling relations. For an easy comparison, the same error for $\MPlanck$ in \autoref{fig:preds_planck} is included as error bars. It is not surprising to see a larger difference between the two results at higher cluster masses as $M_{SZ}$ is normalised to $6\times 10^{14}h_{70}^{-1}\Msun$. It is clear that the blue dotted-dashed line follows the Planck data points very well. While the red dashed line follows the distribution whose mean and $1-sigma$ values are around $0.04^{+0.05}_{-0.05}$ (with a standard error of $\pm 0.0007$) at all mass range which is in agreement with the result from \autoref{fig:preds_simulation}. Although we use the original high-resolution $y$ maps to calculate $Y$, a similar result is obtained using $\mockdata{}$ for resolved clusters. In practice, we do not find any noticeable difference for clusters whose $R_{500}$ is greater than 5 pixels. Therefore, a possible explanation for the fact that $\bP$ is different from 0 lies in the intrinsic difference in the assumed $Y$-$M$ relations between \theth\ simulated clusters and the Planck clusters.

As the $Y-M$ relation imposes the difference between Planck and \theth\ simulation and suggests that the root of the bias shown in Fig. 2 lies in this, we discuss the possible reasons for the differences here. From an observational point of view, the uncertainties may come from a couple of reasons: (1) the calibration of the $Y-M$ relation \citep{ade2014planck} where the cluster mass $M_{500}$ is estimated from X-ray data under the HE assumption. Therefore, an HE mass bias $b_{HE}$ with a value of $\approx 0.1-0.2$ \citep[see][for discussions on this value difference]{giulia300} will inherit in. Furthermore, as presented in \citep{ade2014planck}, the Planck $Y-M$ slope is steeper than several simulations \citep[see Fig. A2 in][]{ade2014planck} which have slopes closer to a self-similar relation. (2) The $Y_{500}$ values derived from the $y$-map are integrated out to $5\times\ R_{500}$ due to the large angular resolution of Planck. The angular resolution impact on the $Y-M$ relation can be found in \cite{Yang2022}. Furthermore, the uncertainty in estimating the $R_{500}$ in observation may also play a role \citep{ferragamo2022comparison}; the mis-center problems \cite{Cui2016} may bias the $Y_{500}$ values as well as $M_{500}$. On simulation side, the uncertainties in the simulated $Y-M$ relation are mainly coming from the implemented baryon models. However, as indicated in \cite{Cui2018,Cui2022}, the same clusters run with three different baryon models, such as GADGET-MUSIC (without AGN feedback), GADGET-X (with AGN feedback) and GIZMO-SIMBA (with strong AGN feedback), show consistent fitting results on the $Y-M$ relation, especially at the massive halo mass end. However, it is worth noting that \cite{Henden2019} showed that including the low mass halo will increase the slope (see references therein for more discussions); meanwhile \cite{Yang2022} also suggested that the angular resolution plays a critical role in this relation. Lastly, though this scaling relation from \theth\ seems almost independent of the implemented gas physics (see also \cite{LeBrun2015}), \cite{LeBrun2017} and \cite{Barnes2017} suggested that different baryon models can violate this self-similarity. Nevertheless, the weak or no redshift evolution of the $Y-M$ relation up to $z = 1$ is generally in agreement with other works (for example, \cite{Henden2019, deAndres2022}).

In addition, it is also worth noting that $\MPlanck$ and $\MCNN$ are intrinsically different: CNN predictions target the true 3D $M_{500}$ based on the physical identified halos in simulation, while $\MPlanck$ is a mass estimated through a calibrated $Y-M$ scaling relation with the integrated $Y$ from observed clusters within $R_{500}$ from 2D images. However, as indicated in \citep[][Fig. A3]{ade2014planck}, the bias is depending on the cluster mass, smaller (0.1) at low cluster mass and larger (0.2) at massive end. This trend is in agreement with the bias  shown in Fig. 2, albeit about 0.1 \% lower (note that $\MPlanck$ used in this work is not bias-corrected). Furthermore, the $Y-M$ relation from \theth\ simulation is in a better agreement with the Planck data at $10^{14}\Msun \lesssim M_{500} \lesssim 4\times10^{14}\Msun$ \citep[see Fig. 10 in][]{Cui2018}. Larger deviation is found at more massive cluster end. Lastly, we also tried cross-model checks with our CNN, i.e. we trained the model with only mock $y-\text{maps}$ of \gadgetx\ and applied it to \simba\ or \gadgetmusic\ mock images (see Supplement G). Our results are qualitatively in agreement with \cite{Villaescusa2021} for a similar approach but to infer cosmological parameters. It suggests that different baryon physics models have a weak impact on our predictions of $M_{500}$ at cluster mass scale. In conclusion, we think that the differences between $\MPlanck$ and $\MCNN$ may mainly result from the $Y-M$ relation. If we trust the $\MCNN$ as the true 3D mass of the clusters, the bias in \cite{ade2014planck} may be just slightly over estimated.






\section{Conclusions}\label{section:conclusions}
CNN is a powerful tool which allows us to directly apply theoretical models or simulation predictions to raw observational data in order to derive quantities that we are interested in.
By training 4 CNNs with mock Planck-like SZ maps and then applying them to real Planck $y$-maps, we evaluate their relevance and provide CNN-estimated masses of the PSZ2 clusters. We use synthetic clusters selected from \theth\ simulation to match the PSZ2 clusters in both redshift and mass ranges. The mock SZ $y$-maps constitute the \cleandata{} sharing the same beam size smoothing as in the real Planck cluster maps. While the \mockdata{} further takes the Planck instrumental noise into account. 4 CNNs are trained independently by separating the full sample ($\sim$200,000 images) into 4 different redshift ranges: $z\leq0.1$, $0.1<z\leq0.2$, $0.2<z\leq0.4$ and $z>0.4$. We show that there are very small biases between the CNN masses and the real 3D cluster masses $M_{500}$ for both \cleandata{} and \mockdata{}, and the scatter in the CNN masses is also very low (an intrinsic scatter -- $16^{th}-84^{th}$ percentile -- of $10\%$ and of $17\%$, respectively; a standard error -- $\pm \sigma/\sqrt{N}$ -- of 0.1\% and of 0.2\%, respectively). By applying these CNNs trained with the \mockdata{} to the \realdata{} cluster maps, we provide newly independent CNN-estimated cluster masses with the posterior uncertainties from the simulation-based inference method. Comparing to the cluster mass estimated by Planck mainly with the HE assumption, we find a relevant non null bias $\bP$ at higher cluster masses, while $\MPlanck$ and $\MCNN$ are in agreement for low mass clusters. After performing an experiment, the fact that the bias between $\MCNN$ and $\MPlanck$ is not zero might be caused by the different slopes of the $Y-M$ scaling law between \theth\ simulations and the Planck one. If the cluster masses estimated by CNN target their true $M_{500}$, this work suggests that the bias for the PSZ2 clusters of $b_p \approx 0.11$ should mostly be due to the HE bias in Planck. This small bias makes the reconciliation with the CMB constraints even harder.

By training CNNs with mock maps and applying them to real cluster maps, this work establishes that ML models can directly link hydrodynamic simulations with observations. Our approach depends less on some theoretical model assumptions and almost does not require estimations on redshift, $R_{500}$, etc. Furthermore, it provides the true simulated analogue physical properties of real observations. To this end, this work is only a starting step towards accurate mass estimations which can potentially be extended to other observations as well.


\section{Methods}\label{section:dataset}

\subsection{THE THREE HUNDRED Simulations}\label{subsec:300sim}

\theth{} project \citep{Cui2018} is based on hydrodynamic zoomed  re-simulations of spherical regions centred on the 324 most massive clusters at $z=0$, identified in the MultiDark dark-matter (DM) only simulation (MDPL2, \cite{Klypin2016}). It utilizes the cosmological parameters from the Planck mission \citep{planck2016}, and simulates a periodic cubic box  of  comoving length $1 \hGpc$ containing $3840^3$ DM particles with mass of $1.5 \times 10^9 \hMsun$. 

A large region around each cluster of $15 \hMpc$ (over 5 times $R_{200}$) is used for re-simulation with different baryonic physics  models: \gadgetmusic \citep{Sembolini2013}, \gadgetx \citep{Murante2010, Rasia2015}, \simba, \cite{Dave2019,Cui2022}). In this study, we mostly focus on the simulated galaxy clusters by \gadgetx.

Halos are identified with the Amiga Halo Finder (AHF)  package \citep{Knollmann2009}. For this work, we select out halos with $M_{500} > 10^{14} h^{-1} M_\odot$ and free of contamination (i.e. the halos do not contain low resolution dark matter particles) out to $z = 1$. The mass and redshift distributions of the selected clusters together with the 1094 PSZ2 clusters are shown in the Supplementary Figure 1. The masses of the Planck clusters are extracted from the PSZ2 catalogue and are further divided (only in the figure) by the expected average hydrostatic bias $b=0.2$ to be compared to the total mass of the synthetic clusters from \theth{} simulations.

For each cluster in the sample, the mock $y$-map is generated with the PYMSZ package \citep{Cui2018,Baldi2018} with 27 different lines of sight projections by rotating the cluster around its centre -- the maximum mass density peak. The Compton-$y$ parameter maps are estimated as follows:
\begin{equation}\label{eq:defy}
    y= \frac{\sigma_{\text{T}}k_{\text{B}}}{m_{\text{e}}c^{2}}\int n_{\text{e}}T_{\text{e}}dl \text{,}
\end{equation}
where $\sigma_{\text{T}}$ is the Thomson cross section, $k_{\text{B}}$ the Boltzmann constant, $c$ the speed of light, $m_{\text{e}}$ the electron rest-mass, $n_{\text{e}}$ the electron number density, $T_{\text{e}}$ is the electron temperature and the integration is done along the observer's line of sight. \autoref{eq:defy} is discretised in our simulated data as in \cite{Sembolini2013} and \cite{LeBrun2015}:
\begin{equation}
    y \simeq \frac{\sigma_{\text{T}}k_{\text{B}}}{m_{\text{e}}c^{2}dA}\sum_{\text{i}}T_{\text{e,i}}N_{\text{e,i}}W(r,h_{i})\text{ ,}
\end{equation}
where $dl$ is substituted by $dV/dA$, $N_{\text{e}}$ is the  electrons density times the volume $V$ and $dA$ is the differential area orthogonal to the line of sight $l$. Moreover, $W(r,h_{\text{i}})$ is the same sph kernel as in the hydrodynamic simulation with smoothing length $h_{\text{i}}$. 

 Originally, each mock image has 1920x1920 pixels with a fixed angular resolution of $5"$ to at least $R_{200}$ in all the clusters. Moreover, the redshift associated with the mock images is the same as the simulation redshift, i.e. the snapshot at which the clusters are selected. In real Planck maps, cluster signals can be affected by foreground or background sources, such as radio sources, sub-millimeter galaxies or other clusters. However, we do not include this contamination into account when generating these mock maps for two reasons: (1) the contamination level is still unclear \citep[see][for different contamination fractions]{Aguado-Barahona2019,Wen2022}; (2) the Y signals from different clusters are indistinguishable with Planck's beam size, especially close to the cluster centre. Therefore, the integrated Y value in Planck might also includes these fore/background clusters, thus, their masses. With these considerations, we do not add contaminating clusters in the mock catalogue, which is a limitation of the current analysis.

\subsection{Mock Planck Observations}
\label{subsec:mockobservations}

In order to apply our trained CNN models to real Planck maps, mock Planck maps must have the same observational limitations, mainly the same angular resolution and noise. The original maps are post-processed using a procedure similar to the one detailed in \citep{Ruppin2019}. We remind it here for the reader's convenience.

Our goal is to create realistic simulations of Planck Compton parameter maps to train our CNN so that it can eventually be applied to cutouts obtained from Gnomonic projections of the publicly available Planck full-sky $y$-map computed with MILCA component separation algorithm. To this end, we need to first smooth the \theth{} $y$-maps (see \textsection \ref{subsec:300sim}) by applying a Gaussian kernel with a 10~arcmin Full Width at Half Maximum (FWHM), filtering the small scales as with the Planck beam. We assume the filtering of large scales by Planck to be negligible. We further process the smoothed simulated $y$-maps by re-gridding them on a grid with 1.7~arcmin pixel resolution in order to match the map resolution of Gnomonic projections of the Planck full-sky $y$-map at HEALPix resolution $\mathrm{Nside}=2048$. This set of maps constitutes what we later call the \cleandata{} in the following. It will be used for characterising the impact of instrumental noise in the CNN predictions.

Then, we generate a full-sky realisation of the Planck instrumental noise based on the publicly available map of the standard deviation of the Compton parameter at HEALPix resolution $\mathrm{Nside}=2048$ and the noise power spectrum of the Planck full-sky $y$-map. Thus, this realisation includes the noise with spatial distribution as observed in the Planck $y$-maps. We extract cutouts of this noise map using Gnomonic projections centred on cluster locations drawn randomly from the PSZ2 catalogue in order to match the noise properties of the detected clusters. These cutouts are generated with the same number of pixels as the maps in the \cleandata{}.

We generate the \mockdata{} by adding a noise map cutout to each map from the \cleandata{}. The maps in this new data set are realistically simulated Planck observations of the synthetic clusters from \theth{} simulation. We note however that we did not include the contamination induced by point sources in these simulated maps. This is examined by using real Planck maps without point-source contamination. As shown in \textsection \ref{section:results}, this should not impact the CNN predictions. We further only select the maps with a higher signal to noise (S/N) ratio, a similar cutoff as in the PSZ2 catalogue. However, this selection is performed with cluster mass cut instead of a S/N limit for two reasons: (1) due to different estimation procedures of the S/N, the S/N for these mock maps shows much larger scatter at lower Planck $S/N$. With a simply S/N cut, even with a higher value, we still found many low mass halos contaminating our sample. Therefore, the corresponding mass cut instead S/N cut gives more reliable maps with higher signal; (2) using $S/N$ cut will produce an uneven separation between training, validation and test much complex. This is because each cluster has 27 random projections and the sample is split by cluster to avoid using the same cluster for training and testing/validating, not by maps.


In summary, the \mockdata{} is composed of the same simulated clusters as the clean mock data set but with the addition of Planck noise. Each simulated data set is composed of 6765 different clusters with 27 rotations amounting to a total of 182,655 maps. Furthermore, these maps are generated from objects extracted from \theth{} simulation to cover the Planck sample PSZ2 in mass $10^{14}M_{\odot}<M<10^{15}M_{\odot}$ and redshift $0<z<1$ which is presented in supplementary Figure 1. Note here that the CNN model is trained and applied in 4 different redshift bins (see Supplement E for the reasons), the selected clusters have a different mass range in each redshift bin. This mass cut is based on our signal-to-noise estimation as presented in previous section. With this mass cut, our sample overlaps with the distribution of Planck clusters well as shown in the supplementary section B (see supplementary Fig.1)

\subsection{Planck full-sky maps}

The work presented in this paper aims at providing new estimates of the total mass of the PSZ2 clusters resulting from the processing of their $y$-maps by a set of trained CNNs. To this end, we extract the map associated with each PSZ2 cluster with known redshift using a Gnomonic projection of the publicly available Planck full-sky $y$-map centred on the galactic coordinates provided in the PSZ2 catalogue. We use the \texttt{gnomview} tool provided by the \texttt{healpy} python library using 96x96 pixels of 1.7x1.7~$arcmin^{2}$ to match the size of the maps in the \mockdata{}. This set of maps forms what we call the \realdata{} in the following.

We further investigate the impact of contamination from astrophysical signal induced from the galactic plane and point sources. We perform the extraction of the PSZ2 cluster $y$-maps again by applying the publicly available masks of the galactic plane and point sources on the full-sky $y$-map before the Gnomonic projections. This allows us to discriminate PSZ2 clusters without any known contamination of the SZ signal within 10~arcmin from the cluster centre defined in the PSZ2 catalogue. In particular, clusters with radio AGN contamination biasing the SZ signal low are excluded with this selection procedure. We find 395 PSZ2 clusters that satisfy this condition. These clusters form what we call the \goldendata{}. We include the Golden sample to verify that our results based on the training with mock maps, which do not consider point sources, are not affected by that.

We present representative examples of $y$-maps from our three data sets with different masses and redshifts in  the supplementary section B (see Fig. 2 of the supplements).




\section*{Data Availability}

The catalogue  of CNN estimated masses for Planck clusters can be downloaded from the following website: \url{https://github.com/The300th/DeepPlanck}

The mock $y$-maps used to train the different CNN models can be accessed upon request to the authors of this paper.

\section*{Code availability}
CNN trained weights are available at \url{https://github.com/The300th/DeepPlanck} together with data products.

\section*{Acknowledgements}

The authors express their sincere thanks to the anonymous referees for their invaluable comments, suggestions and kind help, without which this work would be incomplete. We also acknowledge helpful discussions with Antonio Ferragamo, Federico De Luca and Federico Sembolini.

D.d.A, W.C. and G.Y. thank financial support from Ministerio de Ciencia e Innovación (Spain) under project grant PID2021-122603NB- C21.
W.C. is supported by the STFC AGP Grant ST/V000594/1 and by the Atracción de Talento Contract no. 2020-T1/TIC-19882 granted by the Comunidad de Madrid in Spain. He further acknowledges the science research grants from the China Manned Space Project with NO. CMS-CSST-2021-A01 and CMS-CSST-2021-B01.
M.D.P. acknowledges support from Sapienza Università di Roma thanks to Progetti di Ricerca Medi 2019, RM11916B7540DD8D and Progetti di Ricerca Medi 2020, RM120172B32D5BE2.

\section{Author Contributions}
Daniel de Andres led the project, wrote and run the ML codes and contributed to most of the writing of manuscript. Weiguang Cui developed, ran The300 simulation and prepared the mock observation images with PYMSZ. He also contributed to write most the paper. Florian Ruppin wrote and run the code pipeline to introduce Planck-like limitations into clean mock observations. He also assisted with the writing of the paper. Marco De Petris and Gustavo Yepes assisted with interpretation, manuscript preparation and revision. Giulia Gianfagna, Ichraf Lahouli, Gianmarco Aversano, Romain Dupuis and Mahmoud Jarraya and Jes\'us Vega-Ferrero contributed to this work with the writing of the project and with Machine Learning technicalities.

\section{Competing interests}
The authors declare no competing interests.

\section{references}
\bibliographystyle{naturemag}
\bibliography{example}

\begin{thebibliography}{10}
\expandafter\ifx\csname url\endcsname\relax
  \def\url#1{\texttt{#1}}\fi
\expandafter\ifx\csname urlprefix\endcsname\relax\def\urlprefix{URL }\fi
\providecommand{\bibinfo}[2]{#2}
\providecommand{\eprint}[2][]{\url{#2}}

\bibitem{kravtsov2012formation}
\bibinfo{author}{Kravtsov, A.~V.} \& \bibinfo{author}{Borgani, S.}
\newblock \bibinfo{title}{Formation of galaxy clusters}.
\newblock \emph{\bibinfo{journal}{Annual Review of Astronomy and Astrophysics}}
  \textbf{\bibinfo{volume}{50}}, \bibinfo{pages}{353--409}
  (\bibinfo{year}{2012}).

\bibitem{aghanim2020planck}
\bibinfo{author}{{Planck Collaboration}} \emph{et~al.}
\newblock \bibinfo{title}{{Planck 2018 results. VI. Cosmological parameters}}.
\newblock \emph{\bibinfo{journal}{\aap}} \textbf{\bibinfo{volume}{641}},
  \bibinfo{pages}{A6} (\bibinfo{year}{2020}).
\newblock \eprint{1807.06209}.

\bibitem{Biviano2006}
\bibinfo{author}{{Biviano}, A.} \emph{et~al.}
\newblock \bibinfo{title}{{On the efficiency and reliability of cluster mass
  estimates based on member galaxies}}.
\newblock \emph{\bibinfo{journal}{\aap}} \textbf{\bibinfo{volume}{456}},
  \bibinfo{pages}{23--36} (\bibinfo{year}{2006}).
\newblock \eprint{astro-ph/0605151}.

\bibitem{Becker2011}
\bibinfo{author}{{Becker}, M.~R.} \& \bibinfo{author}{{Kravtsov}, A.~V.}
\newblock \bibinfo{title}{{On the Accuracy of Weak-lensing Cluster Mass
  Reconstructions}}.
\newblock \emph{\bibinfo{journal}{\apj}} \textbf{\bibinfo{volume}{740}},
  \bibinfo{pages}{25} (\bibinfo{year}{2011}).
\newblock \eprint{1011.1681}.

\bibitem{bryan1998}
\bibinfo{author}{{Bryan}, G.~L.} \& \bibinfo{author}{{Norman}, M.~L.}
\newblock \bibinfo{title}{{Statistical Properties of X-Ray Clusters: Analytic
  and Numerical Comparisons}}.
\newblock \emph{\bibinfo{journal}{\apj}} \textbf{\bibinfo{volume}{495}},
  \bibinfo{pages}{80--99} (\bibinfo{year}{1998}).
\newblock \eprint{astro-ph/9710107}.

\bibitem{planck2016b}
\bibinfo{author}{{Planck Collaboration}} \emph{et~al.}
\newblock \bibinfo{title}{{Planck 2015 results. XXIV. Cosmology from
  Sunyaev-Zeldovich cluster counts}}.
\newblock \emph{\bibinfo{journal}{\aap}} \textbf{\bibinfo{volume}{594}},
  \bibinfo{pages}{A24} (\bibinfo{year}{2016}).
\newblock \eprint{1502.01597}.

\bibitem{salvati2018}
\bibinfo{author}{{Salvati}, L.}, \bibinfo{author}{{Douspis}, M.} \&
  \bibinfo{author}{{Aghanim}, N.}
\newblock \bibinfo{title}{{Constraints from thermal Sunyaev-Zel'dovich cluster
  counts and power spectrum combined with CMB}}.
\newblock \emph{\bibinfo{journal}{\aap}} \textbf{\bibinfo{volume}{614}},
  \bibinfo{pages}{A13} (\bibinfo{year}{2018}).
\newblock \eprint{1708.00697}.

\bibitem{giuliabias}
\bibinfo{author}{{Gianfagna}, G.} \emph{et~al.}
\newblock \bibinfo{title}{{Exploring the hydrostatic mass bias in MUSIC
  clusters: application to the NIKA2 mock sample}}.
\newblock \emph{\bibinfo{journal}{\mnras}} \textbf{\bibinfo{volume}{502}},
  \bibinfo{pages}{5115--5133} (\bibinfo{year}{2021}).
\newblock \eprint{2010.03634}.

\bibitem{bishop2006pattern}
\bibinfo{author}{Bishop, C.~M.} \& \bibinfo{author}{Nasrabadi, N.~M.}
\newblock \emph{\bibinfo{title}{Pattern recognition and machine learning}},
  vol.~\bibinfo{volume}{4} (\bibinfo{publisher}{Springer},
  \bibinfo{year}{2006}).

\bibitem{baron2019machine}
\bibinfo{author}{Baron, D.}
\newblock \bibinfo{title}{Machine learning in astronomy: A practical overview}.
\newblock \emph{\bibinfo{journal}{arXiv preprint arXiv:1904.07248}}
  (\bibinfo{year}{2019}).

\bibitem{lecun2015deep}
\bibinfo{author}{LeCun, Y.}, \bibinfo{author}{Bengio, Y.} \&
  \bibinfo{author}{Hinton, G.}
\newblock \bibinfo{title}{Deep learning}.
\newblock \emph{\bibinfo{journal}{nature}} \textbf{\bibinfo{volume}{521}},
  \bibinfo{pages}{436--444} (\bibinfo{year}{2015}).

\bibitem{Goodfellow-et-al-2016}
\bibinfo{author}{Goodfellow, I.}, \bibinfo{author}{Bengio, Y.} \&
  \bibinfo{author}{Courville, A.}
\newblock \emph{\bibinfo{title}{Deep Learning}} (\bibinfo{publisher}{MIT
  Press}, \bibinfo{year}{2016}).
\newblock \bibinfo{note}{\url{http://www.deeplearningbook.org}}.

\bibitem{lecun1989generalization}
\bibinfo{author}{LeCun, Y.} \emph{et~al.}
\newblock \bibinfo{title}{Generalization and network design strategies}.
\newblock \emph{\bibinfo{journal}{Connectionism in perspective}}
  \textbf{\bibinfo{volume}{19}}, \bibinfo{pages}{143--155}
  (\bibinfo{year}{1989}).

\bibitem{ntampaka2019deep}
\bibinfo{author}{Ntampaka, M.} \emph{et~al.}
\newblock \bibinfo{title}{A deep learning approach to galaxy cluster x-ray
  masses}.
\newblock \emph{\bibinfo{journal}{The Astrophysical Journal}}
  \textbf{\bibinfo{volume}{876}}, \bibinfo{pages}{82} (\bibinfo{year}{2019}).

\bibitem{gupta2020massSZ}
\bibinfo{author}{Gupta, N.} \& \bibinfo{author}{Reichardt, C.~L.}
\newblock \bibinfo{title}{Mass estimation of galaxy clusters with deep
  learning. i. sunyaev--zel’dovich effect}.
\newblock \emph{\bibinfo{journal}{The Astrophysical Journal}}
  \textbf{\bibinfo{volume}{900}}, \bibinfo{pages}{110} (\bibinfo{year}{2020}).

\bibitem{gupta2020massWL}
\bibinfo{author}{Gupta, N.} \& \bibinfo{author}{Reichardt, C.}
\newblock \bibinfo{title}{Mass estimation of galaxy clusters with deep learning
  ii. cosmic microwave background cluster lensing}.
\newblock \emph{\bibinfo{journal}{The Astrophysical Journal}}
  \textbf{\bibinfo{volume}{923}}, \bibinfo{pages}{96} (\bibinfo{year}{2021}).

\bibitem{yan2020galaxy}
\bibinfo{author}{Yan, Z.}, \bibinfo{author}{Mead, A.},
  \bibinfo{author}{Van~Waerbeke, L.}, \bibinfo{author}{Hinshaw, G.} \&
  \bibinfo{author}{McCarthy, I.}
\newblock \bibinfo{title}{Galaxy cluster mass estimation with deep learning and
  hydrodynamical simulations}.
\newblock \emph{\bibinfo{journal}{Monthly Notices of the Royal Astronomical
  Society}} \textbf{\bibinfo{volume}{499}}, \bibinfo{pages}{3445--3458}
  (\bibinfo{year}{2020}).

\bibitem{ho2019robust}
\bibinfo{author}{Ho, M.} \emph{et~al.}
\newblock \bibinfo{title}{A robust and efficient deep learning method for
  dynamical mass measurements of galaxy clusters}.
\newblock \emph{\bibinfo{journal}{The Astrophysical Journal}}
  \textbf{\bibinfo{volume}{887}}, \bibinfo{pages}{25} (\bibinfo{year}{2019}).

\bibitem{kodi2021simulation}
\bibinfo{author}{Kodi~Ramanah, D.}, \bibinfo{author}{Wojtak, R.} \&
  \bibinfo{author}{Arendse, N.}
\newblock \bibinfo{title}{Simulation-based inference of dynamical galaxy
  cluster masses with 3d convolutional neural networks}.
\newblock \emph{\bibinfo{journal}{Monthly Notices of the Royal Astronomical
  Society}} \textbf{\bibinfo{volume}{501}}, \bibinfo{pages}{4080--4091}
  (\bibinfo{year}{2021}).

\bibitem{ho2021approximate}
\bibinfo{author}{Ho, M.}, \bibinfo{author}{Farahi, A.}, \bibinfo{author}{Rau,
  M.~M.} \& \bibinfo{author}{Trac, H.}
\newblock \bibinfo{title}{Approximate bayesian uncertainties on deep learning
  dynamical mass estimates of galaxy clusters}.
\newblock \emph{\bibinfo{journal}{The Astrophysical Journal}}
  \textbf{\bibinfo{volume}{908}}, \bibinfo{pages}{204} (\bibinfo{year}{2021}).

\bibitem{planckcatalog}
\bibinfo{author}{{Planck Collaboration}} \emph{et~al.}
\newblock \bibinfo{title}{{Planck 2015 results. XXVII. The second Planck
  catalogue of Sunyaev-Zeldovich sources}}.
\newblock \emph{\bibinfo{journal}{\aap}} \textbf{\bibinfo{volume}{594}},
  \bibinfo{pages}{A27} (\bibinfo{year}{2016}).
\newblock \eprint{1502.01598}.

\bibitem{Cui2018}
\bibinfo{author}{{Cui}, W.} \emph{et~al.}
\newblock \bibinfo{title}{{The Three Hundred project: a large catalogue of
  theoretically modelled galaxy clusters for cosmological and astrophysical
  applications}}.
\newblock \emph{\bibinfo{journal}{\mnras}} \textbf{\bibinfo{volume}{480}},
  \bibinfo{pages}{2898--2915} (\bibinfo{year}{2018}).
\newblock \eprint{1809.04622}.

\bibitem{Planck2013}
\bibinfo{author}{{Planck Collaboration}} \emph{et~al.}
\newblock \bibinfo{title}{Planck 2013 results. xx. cosmology from
  sunyaev-zeldovich cluster counts}.
\newblock \emph{\bibinfo{journal}{A\&A}} \textbf{\bibinfo{volume}{571}},
  \bibinfo{pages}{A20} (\bibinfo{year}{2014}).
\newblock \urlprefix\url{https://doi.org/10.1051/0004-6361/201321521}.

\bibitem{sunyaev1972}
\bibinfo{author}{{Sunyaev}, R.~A.} \& \bibinfo{author}{{Zeldovich}, Y.~B.}
\newblock \bibinfo{title}{{The Observations of Relic Radiation as a Test of the
  Nature of X-Ray Radiation from the Clusters of Galaxies}}.
\newblock \emph{\bibinfo{journal}{Comments on Astrophysics and Space Physics}}
  \textbf{\bibinfo{volume}{4}}, \bibinfo{pages}{173} (\bibinfo{year}{1972}).

\bibitem{ade2014planck}
\bibinfo{author}{{Planck Collaboration}} \emph{et~al.}
\newblock \bibinfo{title}{{Planck 2013 results. XX. Cosmology from
  Sunyaev-Zeldovich cluster counts}}.
\newblock \emph{\bibinfo{journal}{\aap}} \textbf{\bibinfo{volume}{571}},
  \bibinfo{pages}{A20} (\bibinfo{year}{2014}).
\newblock \eprint{1303.5080}.

\bibitem{Kravtsov_2006}
\bibinfo{author}{Kravtsov, A.~V.}, \bibinfo{author}{Vikhlinin, A.} \&
  \bibinfo{author}{Nagai, D.}
\newblock \bibinfo{title}{A new robust low-scatter x-ray mass indicator for
  clusters of galaxies}.
\newblock \emph{\bibinfo{journal}{The Astrophysical Journal}}
  \textbf{\bibinfo{volume}{650}}, \bibinfo{pages}{128--136}
  (\bibinfo{year}{2006}).
\newblock \urlprefix\url{https://doi.org/10.1086/506319}.

\bibitem{Knollmann2009}
\bibinfo{author}{{Knollmann}, S.~R.} \& \bibinfo{author}{{Knebe}, A.}
\newblock \bibinfo{title}{{AHF: Amiga's Halo Finder}}.
\newblock \emph{\bibinfo{journal}{\apjs}} \textbf{\bibinfo{volume}{182}},
  \bibinfo{pages}{608--624} (\bibinfo{year}{2009}).
\newblock \eprint{0904.3662}.

\bibitem{giulia300}
\bibinfo{author}{{Gianfagna}, G.}, \bibinfo{author}{{Rasia}, E.},
  \bibinfo{author}{{Cui}, W.}, \bibinfo{author}{{De Petris}, M.} \&
  \bibinfo{author}{{Yepes}, G.}
\newblock \bibinfo{title}{{The hydrostatic mass bias in The Three Hundred
  clusters}}.
\newblock \emph{\bibinfo{journal}{arXiv e-prints}}
  \bibinfo{pages}{arXiv:2111.01903} (\bibinfo{year}{2021}).
\newblock \eprint{2111.01903}.

\bibitem{Yang2022}
\bibinfo{author}{{Yang}, T.} \emph{et~al.}
\newblock \bibinfo{title}{{Understanding the Sunyaev-Zeldovich decrement versus
  halo mass using the SIMBA and TNG Simulations}}.
\newblock \emph{\bibinfo{journal}{arXiv e-prints}}
  \bibinfo{pages}{arXiv:2202.11430} (\bibinfo{year}{2022}).
\newblock \eprint{2202.11430}.

\bibitem{ferragamo2022comparison}
\bibinfo{author}{Ferragamo, A.} \emph{et~al.}
\newblock \bibinfo{title}{Comparison of hydrostatic and lensing cluster mass
  estimates: A pilot study in macs j0647. 7+ 7015}.
\newblock \emph{\bibinfo{journal}{Astronomy \& Astrophysics}}
  \textbf{\bibinfo{volume}{661}}, \bibinfo{pages}{A65} (\bibinfo{year}{2022}).

\bibitem{Cui2016}
\bibinfo{author}{{Cui}, W.} \emph{et~al.}
\newblock \bibinfo{title}{{nIFTy galaxy cluster simulations - IV. Quantifying
  the influence of baryons on halo properties}}.
\newblock \emph{\bibinfo{journal}{\mnras}} \textbf{\bibinfo{volume}{458}},
  \bibinfo{pages}{4052--4073} (\bibinfo{year}{2016}).
\newblock \eprint{1602.06668}.

\bibitem{Cui2022}
\bibinfo{author}{{Cui}, W.} \emph{et~al.}
\newblock \bibinfo{title}{{THE THREE HUNDRED project: The GIZMO-SIMBA run}}.
\newblock \emph{\bibinfo{journal}{\mnras}} \textbf{\bibinfo{volume}{514}},
  \bibinfo{pages}{977--996} (\bibinfo{year}{2022}).
\newblock \eprint{2202.14038}.

\bibitem{Henden2019}
\bibinfo{author}{{Henden}, N.~A.}, \bibinfo{author}{{Puchwein}, E.} \&
  \bibinfo{author}{{Sijacki}, D.}
\newblock \bibinfo{title}{{The redshift evolution of X-ray and
  Sunyaev-Zel'dovich scaling relations in the FABLE simulations}}.
\newblock \emph{\bibinfo{journal}{\mnras}} \textbf{\bibinfo{volume}{489}},
  \bibinfo{pages}{2439--2470} (\bibinfo{year}{2019}).
\newblock \eprint{1905.00013}.

\bibitem{LeBrun2015}
\bibinfo{author}{{Le Brun}, A. M.~C.}, \bibinfo{author}{{McCarthy}, I.~G.} \&
  \bibinfo{author}{{Melin}, J.-B.}
\newblock \bibinfo{title}{{Testing Sunyaev-Zel'dovich measurements of the hot
  gas content of dark matter haloes using synthetic skies}}.
\newblock \emph{\bibinfo{journal}{\mnras}} \textbf{\bibinfo{volume}{451}},
  \bibinfo{pages}{3868--3881} (\bibinfo{year}{2015}).
\newblock \eprint{1501.05666}.

\bibitem{LeBrun2017}
\bibinfo{author}{{Le Brun}, A. M.~C.}, \bibinfo{author}{{McCarthy}, I.~G.},
  \bibinfo{author}{{Schaye}, J.} \& \bibinfo{author}{{Ponman}, T.~J.}
\newblock \bibinfo{title}{{The scatter and evolution of the global hot gas
  properties of simulated galaxy cluster populations}}.
\newblock \emph{\bibinfo{journal}{\mnras}} \textbf{\bibinfo{volume}{466}},
  \bibinfo{pages}{4442--4469} (\bibinfo{year}{2017}).
\newblock \eprint{1606.04545}.

\bibitem{Barnes2017}
\bibinfo{author}{{Barnes}, D.~J.} \emph{et~al.}
\newblock \bibinfo{title}{{The redshift evolution of massive galaxy clusters in
  the MACSIS simulations}}.
\newblock \emph{\bibinfo{journal}{\mnras}} \textbf{\bibinfo{volume}{465}},
  \bibinfo{pages}{213--233} (\bibinfo{year}{2017}).
\newblock \eprint{1607.04569}.

\bibitem{deAndres2022}
\bibinfo{author}{{de Andres}, D.} \emph{et~al.}
\newblock \bibinfo{title}{{Machine Learning methods to estimate observational
  properties of galaxy clusters in large volume cosmological N-body
  simulations}}.
\newblock \emph{\bibinfo{journal}{arXiv e-prints}}
  \bibinfo{pages}{arXiv:2204.10751} (\bibinfo{year}{2022}).
\newblock \eprint{2204.10751}.

\bibitem{Villaescusa2021}
\bibinfo{author}{Villaescusa-Navarro, F.} \emph{et~al.}
\newblock \bibinfo{title}{Robust marginalization of baryonic effects for
  cosmological inference at the field level}.
\newblock \emph{\bibinfo{journal}{arXiv preprint arXiv:2109.10360}}
  (\bibinfo{year}{2021}).

\bibitem{Klypin2016}
\bibinfo{author}{{Klypin}, A.}, \bibinfo{author}{{Yepes}, G.},
  \bibinfo{author}{{Gottl{\"o}ber}, S.}, \bibinfo{author}{{Prada}, F.} \&
  \bibinfo{author}{{He{\ss}}, S.}
\newblock \bibinfo{title}{{MultiDark simulations: the story of dark matter halo
  concentrations and density profiles}}.
\newblock \emph{\bibinfo{journal}{\mnras}} \textbf{\bibinfo{volume}{457}},
  \bibinfo{pages}{4340--4359} (\bibinfo{year}{2016}).
\newblock \eprint{1411.4001}.

\bibitem{planck2016}
\bibinfo{author}{{Planck Collaboration}} \emph{et~al.}
\newblock \bibinfo{title}{{Planck 2015 results. XIII. Cosmological
  parameters}}.
\newblock \emph{\bibinfo{journal}{\aap}} \textbf{\bibinfo{volume}{594}},
  \bibinfo{pages}{A13} (\bibinfo{year}{2016}).
\newblock \eprint{1502.01589}.

\bibitem{Sembolini2013}
\bibinfo{author}{{Sembolini}, F.} \emph{et~al.}
\newblock \bibinfo{title}{{The MUSIC of galaxy clusters - I. Baryon properties
  and scaling relations of the thermal Sunyaev-Zel'dovich effect}}.
\newblock \emph{\bibinfo{journal}{\mnras}} \textbf{\bibinfo{volume}{429}},
  \bibinfo{pages}{323--343} (\bibinfo{year}{2013}).
\newblock \eprint{1207.4438}.

\bibitem{Murante2010}
\bibinfo{author}{{Murante}, G.}, \bibinfo{author}{{Monaco}, P.},
  \bibinfo{author}{{Giovalli}, M.}, \bibinfo{author}{{Borgani}, S.} \&
  \bibinfo{author}{{Diaferio}, A.}
\newblock \bibinfo{title}{{A subresolution multiphase interstellar medium model
  of star formation and supernova energy feedback}}.
\newblock \emph{\bibinfo{journal}{\mnras}} \textbf{\bibinfo{volume}{405}},
  \bibinfo{pages}{1491--1512} (\bibinfo{year}{2010}).
\newblock \eprint{1002.4122}.

\bibitem{Rasia2015}
\bibinfo{author}{{Rasia}, E.} \emph{et~al.}
\newblock \bibinfo{title}{{Cool Core Clusters from Cosmological Simulations}}.
\newblock \emph{\bibinfo{journal}{\apjl}} \textbf{\bibinfo{volume}{813}},
  \bibinfo{pages}{L17} (\bibinfo{year}{2015}).
\newblock \eprint{1509.04247}.

\bibitem{Dave2019}
\bibinfo{author}{{Dav{\'e}}, R.} \emph{et~al.}
\newblock \bibinfo{title}{{SIMBA: Cosmological simulations with black hole
  growth and feedback}}.
\newblock \emph{\bibinfo{journal}{\mnras}} \textbf{\bibinfo{volume}{486}},
  \bibinfo{pages}{2827--2849} (\bibinfo{year}{2019}).
\newblock \eprint{1901.10203}.

\bibitem{Baldi2018}
\bibinfo{author}{{Baldi}, A.~S.} \emph{et~al.}
\newblock \bibinfo{title}{{Kinetic Sunyaev-Zel'dovich effect in rotating galaxy
  clusters from MUSIC simulations}}.
\newblock \emph{\bibinfo{journal}{\mnras}} \textbf{\bibinfo{volume}{479}},
  \bibinfo{pages}{4028--4040} (\bibinfo{year}{2018}).
\newblock \eprint{1805.07142}.

\bibitem{Aguado-Barahona2019}
\bibinfo{author}{{Aguado-Barahona}, A.} \emph{et~al.}
\newblock \bibinfo{title}{{Optical validation and characterization of Planck
  PSZ2 sources at the Canary Islands observatories. II. Second year of LP15
  observations}}.
\newblock \emph{\bibinfo{journal}{\aap}} \textbf{\bibinfo{volume}{631}},
  \bibinfo{pages}{A148} (\bibinfo{year}{2019}).
\newblock \eprint{1909.06235}.

\bibitem{Wen2022}
\bibinfo{author}{{Wen}, Z.~L.} \& \bibinfo{author}{{Han}, J.~L.}
\newblock \bibinfo{title}{{Clusters of galaxies up to z = 1.5 identified from
  photometric data of the Dark Energy Survey and unWISE}}.
\newblock \emph{\bibinfo{journal}{\mnras}}  (\bibinfo{year}{2022}).
\newblock \eprint{2204.11215}.

\bibitem{Ruppin2019}
\bibinfo{author}{Ruppin, F.} \emph{et~al.}
\newblock \bibinfo{title}{{Impact of ICM disturbances on the mean pressure
  profile of galaxy clusters: a prospective study of the NIKA2 SZ large program
  with MUSIC synthetic clusters}}.
\newblock \emph{\bibinfo{journal}{Astron. Astrophys.}}
  \textbf{\bibinfo{volume}{631}}, \bibinfo{pages}{A21} (\bibinfo{year}{2019}).
\newblock \eprint{1901.04580}.

\bibitem{simonyan2014very}
\bibinfo{author}{Simonyan, K.} \& \bibinfo{author}{Zisserman, A.}
\newblock \bibinfo{title}{Very deep convolutional networks for large-scale
  image recognition}.
\newblock \emph{\bibinfo{journal}{arXiv preprint arXiv:1409.1556}}
  (\bibinfo{year}{2014}).

\bibitem{riesenhuber1999hierarchical}
\bibinfo{author}{Riesenhuber, M.} \& \bibinfo{author}{Poggio, T.}
\newblock \bibinfo{title}{Hierarchical models of object recognition in cortex}.
\newblock \emph{\bibinfo{journal}{Nature neuroscience}}
  \textbf{\bibinfo{volume}{2}}, \bibinfo{pages}{1019--1025}
  (\bibinfo{year}{1999}).

\bibitem{nair2010rectified}
\bibinfo{author}{Nair, V.} \& \bibinfo{author}{Hinton, G.~E.}
\newblock \bibinfo{title}{Rectified linear units improve restricted boltzmann
  machines}.
\newblock In \emph{\bibinfo{booktitle}{Icml}} (\bibinfo{year}{2010}).

\bibitem{srivastava2014dropout}
\bibinfo{author}{Srivastava, N.}, \bibinfo{author}{Hinton, G.},
  \bibinfo{author}{Krizhevsky, A.}, \bibinfo{author}{Sutskever, I.} \&
  \bibinfo{author}{Salakhutdinov, R.}
\newblock \bibinfo{title}{Dropout: a simple way to prevent neural networks from
  overfitting}.
\newblock \emph{\bibinfo{journal}{The journal of machine learning research}}
  \textbf{\bibinfo{volume}{15}}, \bibinfo{pages}{1929--1958}
  (\bibinfo{year}{2014}).

\bibitem{resnet}
\bibinfo{author}{He, K.}, \bibinfo{author}{Zhang, X.}, \bibinfo{author}{Ren,
  S.} \& \bibinfo{author}{Sun, J.}
\newblock \bibinfo{title}{Deep residual learning for image recognition}.
\newblock In \emph{\bibinfo{booktitle}{Proceedings of the IEEE conference on
  computer vision and pattern recognition}}, \bibinfo{pages}{770--778}
  (\bibinfo{year}{2016}).

\bibitem{inception}
\bibinfo{author}{Szegedy, C.} \emph{et~al.}
\newblock \bibinfo{title}{Going deeper with convolutions}.
\newblock In \emph{\bibinfo{booktitle}{Proceedings of the IEEE conference on
  computer vision and pattern recognition}}, \bibinfo{pages}{1--9}
  (\bibinfo{year}{2015}).

\bibitem{xception}
\bibinfo{author}{Chollet, F.}
\newblock \bibinfo{title}{Xception: Deep learning with depthwise separable
  convolutions}.
\newblock In \emph{\bibinfo{booktitle}{Proceedings of the IEEE conference on
  computer vision and pattern recognition}}, \bibinfo{pages}{1251--1258}
  (\bibinfo{year}{2017}).

\bibitem{kingma2014adam}
\bibinfo{author}{Kingma, D.~P.} \& \bibinfo{author}{Ba, J.}
\newblock \bibinfo{title}{Adam: A method for stochastic optimization}.
\newblock \emph{\bibinfo{journal}{arXiv preprint arXiv:1412.6980}}
  (\bibinfo{year}{2014}).

\bibitem{chollet2015keras}
\bibinfo{author}{Chollet, F.} \emph{et~al.}
\newblock \bibinfo{title}{Keras}.
\newblock \bibinfo{howpublished}{\url{https://keras.io}}
  (\bibinfo{year}{2015}).

\bibitem{abadi2016tensorflow}
\bibinfo{author}{Abadi, M.} \emph{et~al.}
\newblock \bibinfo{title}{Tensorflow: Large-scale machine learning on
  heterogeneous distributed systems}.
\newblock \emph{\bibinfo{journal}{arXiv preprint arXiv:1603.04467}}
  (\bibinfo{year}{2016}).

\bibitem{SPTcatalog}
\bibinfo{author}{Bleem, L.~E.} \emph{et~al.}
\newblock \bibinfo{title}{The {SPTpol} extended cluster survey}.
\newblock \emph{\bibinfo{journal}{The Astrophysical Journal Supplement Series}}
  \textbf{\bibinfo{volume}{247}}, \bibinfo{pages}{25} (\bibinfo{year}{2020}).
\newblock \urlprefix\url{https://doi.org/10.3847/1538-4365/ab6993}.

\bibitem{bleem2015galaxy}
\bibinfo{author}{Bleem, L.} \emph{et~al.}
\newblock \bibinfo{title}{Galaxy clusters discovered via the
  sunyaev--zel'dovich effect in the 2500-square-degree spt-sz survey}.
\newblock \emph{\bibinfo{journal}{The Astrophysical Journal Supplement Series}}
  \textbf{\bibinfo{volume}{216}}, \bibinfo{pages}{27} (\bibinfo{year}{2015}).

\bibitem{Postman2012}
\bibinfo{author}{{Postman}, M.} \emph{et~al.}
\newblock \bibinfo{title}{{The Cluster Lensing and Supernova Survey with
  Hubble: An Overview}}.
\newblock \emph{\bibinfo{journal}{\apjs}} \textbf{\bibinfo{volume}{199}},
  \bibinfo{pages}{25} (\bibinfo{year}{2012}).
\newblock \eprint{1106.3328}.

\bibitem{Hoekstra2013}
\bibinfo{author}{{Hoekstra}, H.} \emph{et~al.}
\newblock \bibinfo{title}{{Masses of Galaxy Clusters from Gravitational
  Lensing}}.
\newblock \emph{\bibinfo{journal}{\ssr}} \textbf{\bibinfo{volume}{177}},
  \bibinfo{pages}{75--118} (\bibinfo{year}{2013}).
\newblock \eprint{1303.3274}.

\bibitem{Andreon2016}
\bibinfo{author}{{Andreon}, S.}
\newblock \bibinfo{title}{{Richness-based masses of rich and famous galaxy
  clusters}}.
\newblock \emph{\bibinfo{journal}{\aap}} \textbf{\bibinfo{volume}{587}},
  \bibinfo{pages}{A158} (\bibinfo{year}{2016}).
\newblock \eprint{1601.06912}.

\bibitem{Oguri2014}
\bibinfo{author}{{Oguri}, M.}
\newblock \bibinfo{title}{{A cluster finding algorithm based on the multiband
  identification of red sequence galaxies}}.
\newblock \emph{\bibinfo{journal}{\mnras}} \textbf{\bibinfo{volume}{444}},
  \bibinfo{pages}{147--161} (\bibinfo{year}{2014}).
\newblock \eprint{1407.4693}.

\bibitem{Andreon2017}
\bibinfo{author}{{Andreon}, S.}, \bibinfo{author}{{Trinchieri}, G.},
  \bibinfo{author}{{Moretti}, A.} \& \bibinfo{author}{{Wang}, J.}
\newblock \bibinfo{title}{{Intrinsic scatter of caustic masses and hydrostatic
  bias: An observational study}}.
\newblock \emph{\bibinfo{journal}{\aap}} \textbf{\bibinfo{volume}{606}},
  \bibinfo{pages}{A25} (\bibinfo{year}{2017}).
\newblock \eprint{1706.08353}.

\bibitem{charnock2018SBI}
\bibinfo{author}{Charnock, T.}, \bibinfo{author}{Lavaux, G.} \&
  \bibinfo{author}{Wandelt, B.~D.}
\newblock \bibinfo{title}{Automatic physical inference with information
  maximizing neural networks}.
\newblock \emph{\bibinfo{journal}{Phys. Rev. D}} \textbf{\bibinfo{volume}{97}},
  \bibinfo{pages}{083004} (\bibinfo{year}{2018}).
\newblock \urlprefix\url{https://link.aps.org/doi/10.1103/PhysRevD.97.083004}.

\bibitem{DeepDream}
\bibinfo{author}{Alexander~Mordvintsev, M.~T., Christopher~Olah}.
\newblock \bibinfo{title}{Deep dream}.
\newblock \emph{\bibinfo{journal}{Google AI Blog}}  (\bibinfo{year}{2015}).

\bibitem{Kim2017}
\bibinfo{author}{{Kim}, B.} \emph{et~al.}
\newblock \bibinfo{title}{{Interpretability Beyond Feature Attribution:
  Quantitative Testing with Concept Activation Vectors (TCAV)}}.
\newblock \emph{\bibinfo{journal}{arXiv e-prints}}
  \bibinfo{pages}{arXiv:1711.11279} (\bibinfo{year}{2017}).
\newblock \eprint{1711.11279}.

\end{thebibliography}
\end{multicols}
\newpage
\section{Supplementary information}


\appendix


\setcounter{figure}{0}   
\setcounter{equation}{0}   
\setcounter{table}{0}   
\renewcommand{\figurename}{Supp. Figure}
\renewcommand{\tablename}{Supp. Table}

\renewcommand{\theequation}{S.\arabic{equation}}

\renewcommand{\figureautorefname}{Supp. Figure}
\renewcommand{\tableautorefname}{Supp. Table}

\section{Notation}
\label{appendix-A}
In this Appendix, we describe the notation used throughout this paper regarding the different masses considered:
\begin{itemize}
    \item $\Mtrue$: 3D dynamical mass of the simulated cluster.
    \item $\MCNN$: The predicted mass by our CNN model. 
    \item $\MSZ$: The mass estimated using the Equation (4) of the main article.
    \item $\MPlanck$: The mass provided in the PSZ2 catalogue by \cite{planckcatalog}.
    
Moreover, all logarithmic values considered are the decimal logarithm, i.e. $\log x = \log_{10} x$.
\end{itemize}

\section{Data set: Mass-redshift distributions and examples of $y$-maps.}
\begin{figure}
\includegraphics[width=\textwidth]{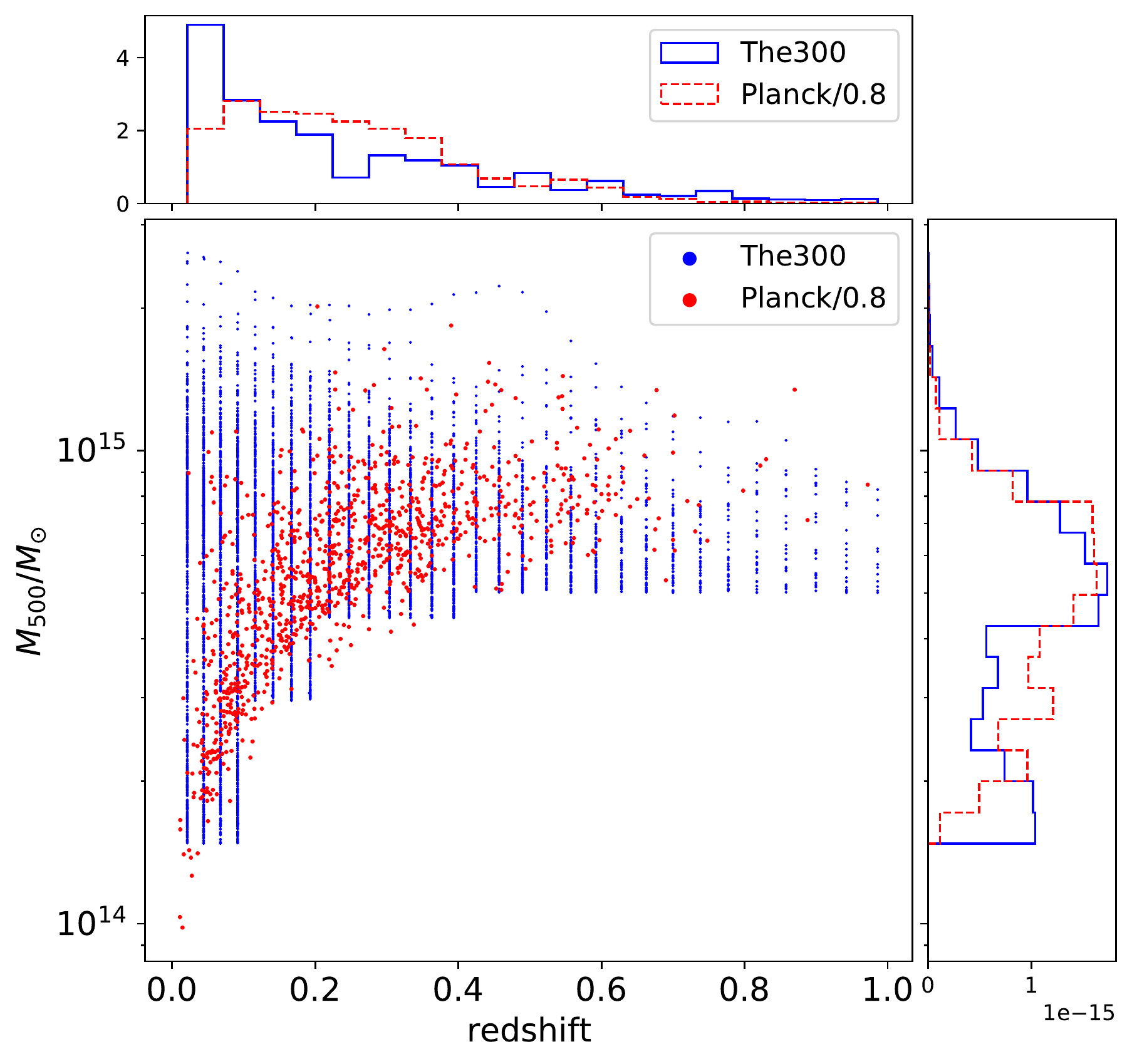}
\caption{{\bf Distributions of mock clusters and real PSZ2 clusters.} Cluster mass, $M_{500}$, distribution along the redshift for the selected clusters from \theth{} \mockdata{} (blue) and \realdata{} PSZ2 catalogue (red). Note that for \theth\ data we show the 3D-dynamical total mass $M_{\text{true}}$. Nevertheless, Planck masses are divided by 0.8 to account for their reported mean hydrostatic mass bias. In the marginal plots, the normalised distributions are shown.}
\label{fig:Mvsz}
\end{figure}

\begin{figure*}
\includegraphics[width=0.9\textwidth]{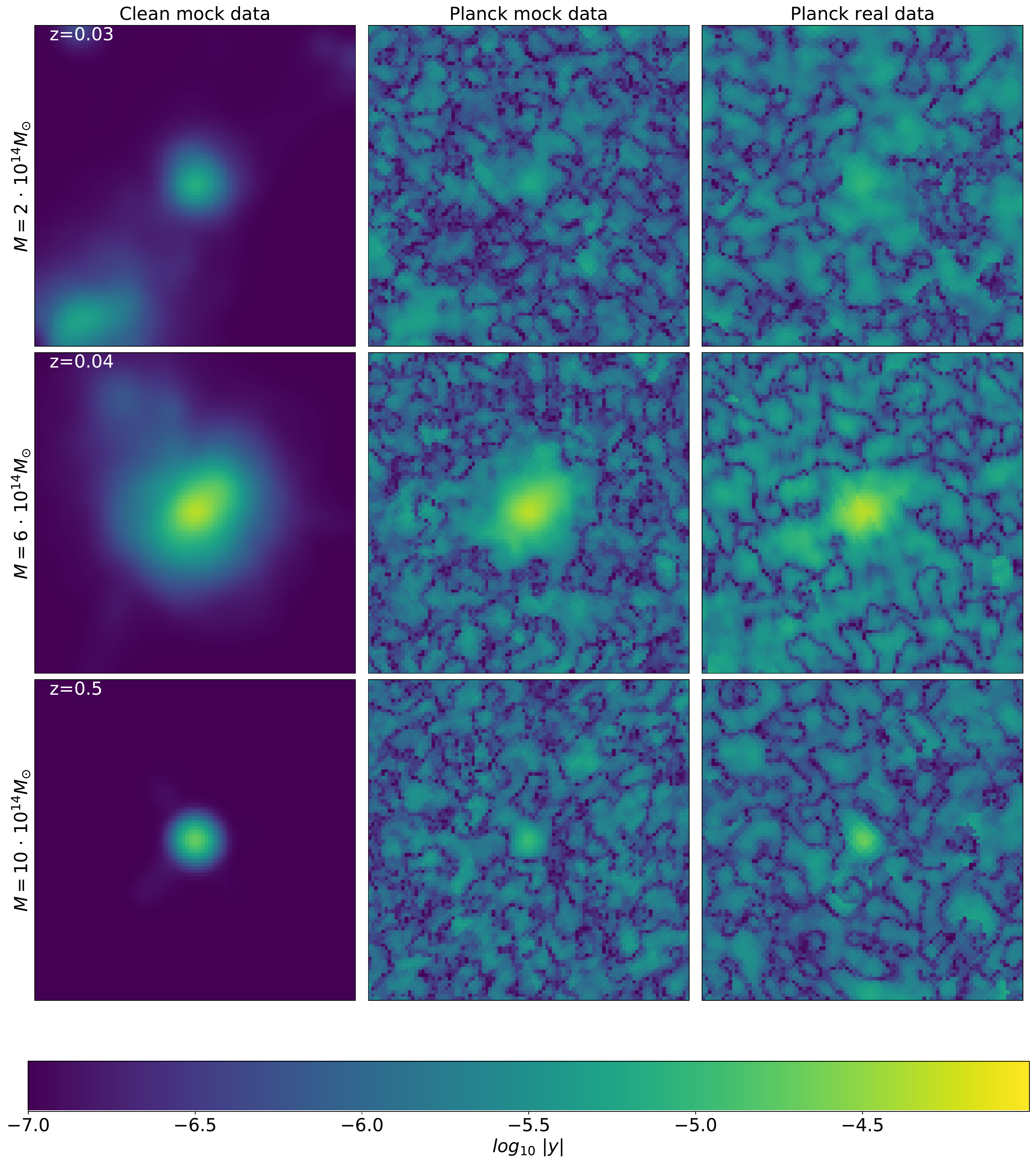}
\caption{{\bf Examples of mock $y-$maps in different data sets.} The SZ maps of selected clusters corresponding to \theth{} and the Planck data for different masses and redshifts (rows). The first column represents \cleandata{}, the second column \mockdata{} and the third column \realdata{}. The first two rows show two nearby clusters, i.e. $z<0.1$, while the third row is for a massive ($10^{15}\Msun$), high redshift $z=0.5$, cluster. The size of the maps is 96x96 pixels and one pixel corresponds to 1.7x1.7 $\text{arcmin}^{2}$.}
\label{fig:maps}
\end{figure*}

In this section, we provide the distributions of mock clusters and real PSZ2 clusters (\autoref{fig:Mvsz}) in mass and redshift and also examples of \cleandata{}, \mockdata{} and \realdata{} clusters (\autoref{fig:maps}). This section supplements the method section in the main article. 

\section{Deep learning model}
\label{appendix-B}
In this Appendix, we first describe the considered deep learning model and then explain how it has been trained and the validation procedure.
\subsection{The CNN}
A feedforward neural networks or MLP defines a mapping 
\begin{equation}
    \text{MLP}(x,w)=y,
\end{equation}
and learns the value of the parameters (weights) $w$ that best fit the equation to some known data $x,y$. This model is called feedforward because information flows from the input $x$ to the intermediate neurons defined in $\text{MLP}(x,w)$ and reaches the final output value $y$. Therefore, the $i$th hidden layer propagates the information forward using an activation function $g$ on top of an affine transformation
\begin{equation}
    \textbf{h}^{i+1}=g\left((\textbf{W}^{T})^{i+1}\textbf{h}^{i}+\textbf{b}^{i+1}\right) ,
\end{equation}
where $\textbf{h}^{i+1}$ is the output (vector) proceeding the $\textbf{h}^{i}$ hidden layer.  $\textbf{W}^{i}$ is a matrix of weights $w$ and $\textbf{b}^{i}$ is a vector of biases. Neural networks are trained by defining a loss function $\mathcal{L}(w)$ which has to be minimised with respect to the weights $w$. In most cases, the minimisation procedure is equivalent to using  maximum likelihood estimation.

Furthermore, a convolution operation is defined to account for sparse interactions, parameter sharing and equivariance to translation. However, convolutions are not naturally equivariant to some other transformations, such us scaling or rotations and thus, other mechanisms are needed in order to handle these transformations. One such mechanism is based on data augmentation, e.g. several rotations of the same image are given to the network aiming at, not only increasing the size of the training set, but also imposing symmetry under this particular transformation. The convolution operation is defined as
\begin{equation}
    (x*w)(t)=\int x(a)w(t-a)da,
\end{equation}
which when applied to a 2D image $I(i,j)$ with a 2D kernel $K$ reads:
\begin{equation}
    (I*K)(i,j) = \sum_{m}\sum_{n}I(m,n)K(i-m,j-n).
\end{equation}

In this work, we use CNNs for predicting the total mass of galaxy clusters inside $R_{500}$ directly from observed SZ maps which represent a two dimensional image $I_{ij}$, i.e. :
\begin{equation}
    \text{CNN}(I_{ij})=M_{500}.
\end{equation}

\begin{figure}
\includegraphics[width=\columnwidth]{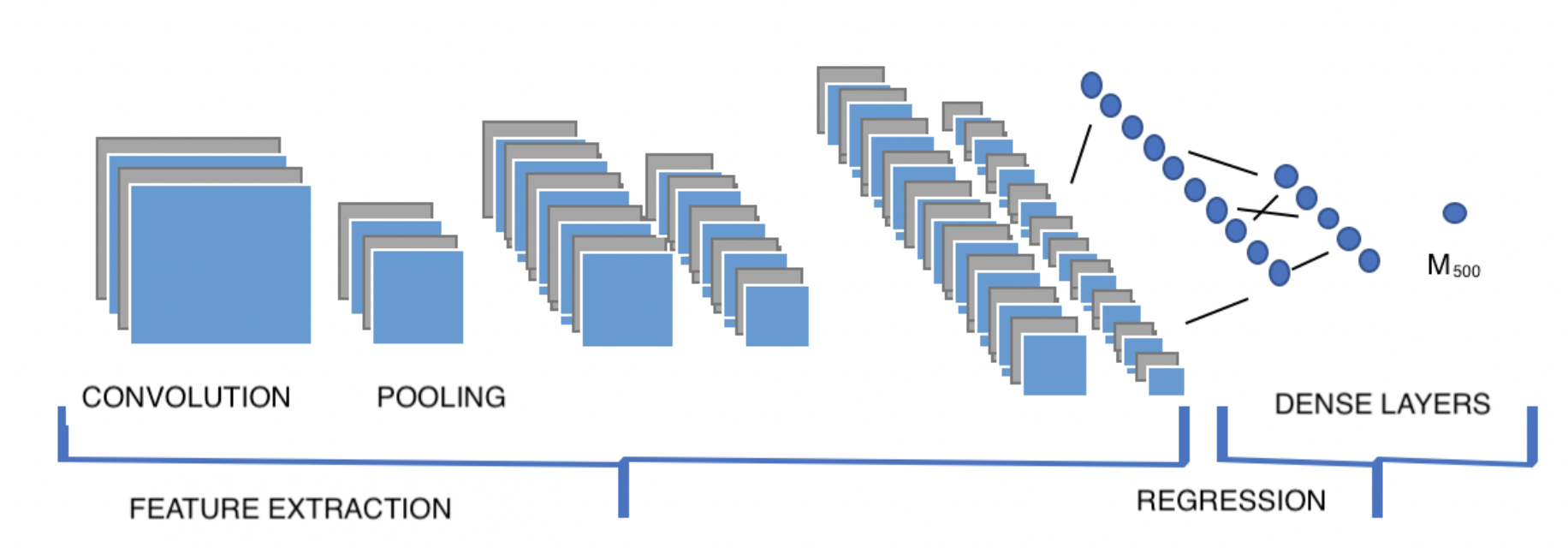}

\caption{CNN architecture. A sequence of convolutional and pooling layers is used for feature extraction. Then, fully connected dense layers are in charge of the regression task in order to obtain $M_{500}$.}
\label{fig:architecture}
\end{figure}

Particularly, we use a convolutional neural network for large-scale image recognition based on the work of \cite{simonyan2014very} known as VGGNet.  This architecture has been successfully applied to infer cluster masses corresponding to different simulated observational maps in \cite{ntampaka2019deep} and \cite{yan2020galaxy}. Here, we apply a similar architecture: 

\begin{description}
    \item 1. 3x3 convolution with 16 filters
    \item 2. 2x2 stride-2 max pooling
    \item 3. 3x3 convolution with 32 filters
    \item 4. 2x2, stride-2 max pooling
    \item 5. 3x3 convolution with 64 filters
    \item 6. 2x2 stride-2 max pooling
    \item 7. global average pooling
    \item 8. 10\% dropout
    \item 9. 200 neurons, dense fully connected
    \item 10. 10\% dropout
    \item 11. 100 neurons, dense  fully connected
    \item 12. 20 neurons, dense fully connected
    \item 13. output neuron
\end{description}
This architecture uses first three pairs of convolutional and pooling layers for feature extraction \citep{riesenhuber1999hierarchical} . Then, it makes use of dense fully connected layers to find a regression between the extracted features and the total mass of the cluster as illustrated in figure \ref{fig:architecture}. The activation  function of these layers is  the rectified linear unit (ReLU, \cite{nair2010rectified}) and dropout is used to avoid overfitting \citep{srivastava2014dropout}. We have checked that increasing the number of neurons in the regression part only yields to overfitting or similar results. Modifying the parameters in the feature extraction (convolutional) part and adding more layers does not improve the performance on the validation set either. Therefore, a VGGNet-based architecture  will difficultly improve the performance and more complex models might be needed to be taken into consideration such as ResNet \citep{resnet}, Inception \citep{inception}, and Xception \citep{xception} networks. Another possible reason behind this is that our problem is not highly complex and the architecture used in this work might be sufficient for analysing our data set.

In order to fit the model, we used the Adam Optimizer \citep{kingma2014adam} with learning rate $10^{-4}$ and  the logarithmic mean squared error as our loss function $\mathcal{L}$, i.e. 
\begin{equation}\label{eq:lossf}
    \mathcal{L}=\frac{1}{n}\sum_{i=1}^{n}(\log \Mtrue^{i} -\log \MCNN^{i})^{2}\text{ ,}
\end{equation}
where $\Mtrue$ is the 3D dynamical ``true" mass and $\MCNN$ the predicted mass. Moreover, we computed $\Mtrue$ by summing over all gravitationally bounded particles inside $R_{500}$, i.e. $M_{\text{true}}=M_{500,\text{true}}$. 

\subsection{Training and Validation}

Before training, the images are normalised as
\begin{equation}
    \hat{X}_{i}=\frac{(X_{i}-\text{mean}(X))}{\text{std}(X)}\text{ ,}
\end{equation}
where $X$ denotes all the maps and $X_{i}$ one image and therefore, $\text{mean}(X)$ is the mean and $\text{std}(X)$ the standard deviation over the whole training set and all their pixel values. Given the randomness of our training procedure, the CNNs converge to different local minimum when fitting the weights for different random initialisation. To study the randomness of this process, we train 100 CNNs (The same architecture is used for all the CNNs) and discriminate the best among them through 100 runs where we split our data set in 80\% training, 10\% validation and 10\% test. The test set is the same for the 100 runs and the train and validation sets are randomly selected from the remaining 90\% of our data. Moreover, we split our data taking into account that a cluster is not used twice for training, validation or testing. Once the models are trained, only the test set is used to select the best model according to the following criteria:

\begin{enumerate}
    \item We compute the relative error as a function of the predicted mass, $\text{error}(\MCNN)=(\MCNN-\Mtrue)/\MCNN$. This error shows how the predictions deviates from the ``true" mass for a given particular predicted mass $\MCNN$. 
    \item  This error is then binned as a function of $\MCNN$ using ten bins containing roughly the same number of images and we compute the mean $\mu_{\text{bin}}$ and standard deviation $\sigma_{\text{bin}}$ per bin. 
    \item Finally, among all the CNNs where $\mu_{\text{bin}}<0.05$ for all the bins, we select the network whose standard deviation is the least, i.e. $\sum_{\text{bin}}\sigma_{\text{bin}}$ is minimum using the validation set. Note that the final training result, i.e Figure 1 in the main article, is shown using only the test set.
\end{enumerate}

Furthermore, four CNNs are trained for these redshift intervals: $0<z_{1}\leq0.1$, $0.1<z_{2}\leq0.2$, $0.2<z_{3}\leq0.4$, $0.4<z_{4}\leq1$. Therefore, we need the redshift of the observed clusters as ancillary data. We refer to the section \ref{sec:redbin} in the supplements for the reason of this choice of number of redshift bins.

\begin{figure}
\includegraphics[width=\columnwidth]{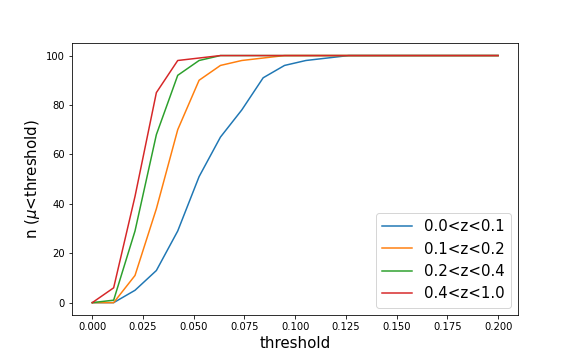}

\caption{The number of models whose $\mu_{\text{bin}}<\text{threshold}$ for all bins is shown. Different colours represent the different redshifts intervals.}
\label{fig:threshold}
\end{figure}

The validation performance can be seen in \autoref{fig:threshold}, where we show the number of models $n$ whose $\mu_{\text{bin}}<\text{threshold}$ for all the bins. We can see that the number of unbiased models (according to our validation criteria) increases with respect to the particular threshold value and the redshift interval. Specifically, we have used the threshold value of $\mu_{\text{bin}}<0.05$ where there are over 40 models that satisfy our criteria.


As far as software is concerned, we have only used open source libraries: the Keras library \citep{chollet2015keras} with Tensorflow  \citep{abadi2016tensorflow} backend. Moreover, we trained our algorithm using one NVIDIA A100 and training on 51408 maps takes 9 seconds per epoch with batch size of 32 images. Note that the number of maps in the training sets varies according to the selected redshift interval. We train our algorithm 200 epochs and we select the model at the epoch at which the validation loss is minimum, not the train loss. The validation loss function usually reaches a minimum in 50 epochs. However, it converges to different local minima. This technique is similar to `early stopping' in the sense that not the last epoch is the best epoch and the hyperparameter `epoch' is fitted using the validation set.


There are two uncertainties: aleatoric and epistemic, which are important in CNN models. The first one mostly correlates with the input-output data. With insufficient input data, the true outputs can be not precisely estimated, even with an ideal model. While the second strongly relates to flexibility in the model. Limited models, such as insufficient network depth, training time, or training catalogue diversity, may not be able to tightly constrain the optimal model parameters. Therefore, we test different network architectures here. In order to do this, we define a convolutional layer as:
\begin{itemize}
    \item 2D Convolution (`kernel\_size',`filters')
    \item MaxPooling2D(pool\_size=(2,2)))
\end{itemize}
where `kernel\_size',`filters',`dropout' are hyperparameters. We also define a dense layer as:
\begin{itemize}
    \item Dense (`number of neurons')
    \item `dropout'
\end{itemize}
here `number of neurons' and `dropout' are hyperparameters. With this two different layers we can define a network architecture as:
\begin{itemize}
    \item Convolutional layer $\times$ `number\_of\_convolutional\_layers'
    \item Global average pooling
    \item `dropout'
    \item Dense layer Layer $\times$ `number\_of\_dense\_layers'
    \item `Dense(100 neurons)'
    \item `Dense(20 neurons)'
    \item `output mass'
\end{itemize}
We further check that any k-subsequent convolutional layer has `filters' $\times2^{k}$ number of filters and any subsequent  k-dense layer has the same number of neurons. We train 1 model (only 1 run per hyperparameter values) on the same validation set with the following hyperparameters:\{`n\_conv\_layers':[3,4,5],
        `n\_dense\_layers':[1,2],
        `number\_of\_neurons':[200]
        `kernel\_size':[1,2,3],
        `filters':[16,32],
        `dropout':[0.1,0.2]\}.
Note that for \{`n\_conv\_layers':[3],
        `n\_dense\_layers':[1],
        `number\_of\_neurons':[200]
        `kernel\_size':[3],
        `filters':[16],
        `dropout':[0.1]\}
we recover our base model used in this work. Lastly, we also test different values of the learning rate from $10^{-2}$ and $10^{-5}$ and determine by visual inspection that $10^{-4}$ is an appropriate value and the training loss converges. 

\begin{figure}
\includegraphics[width=1.\textwidth]{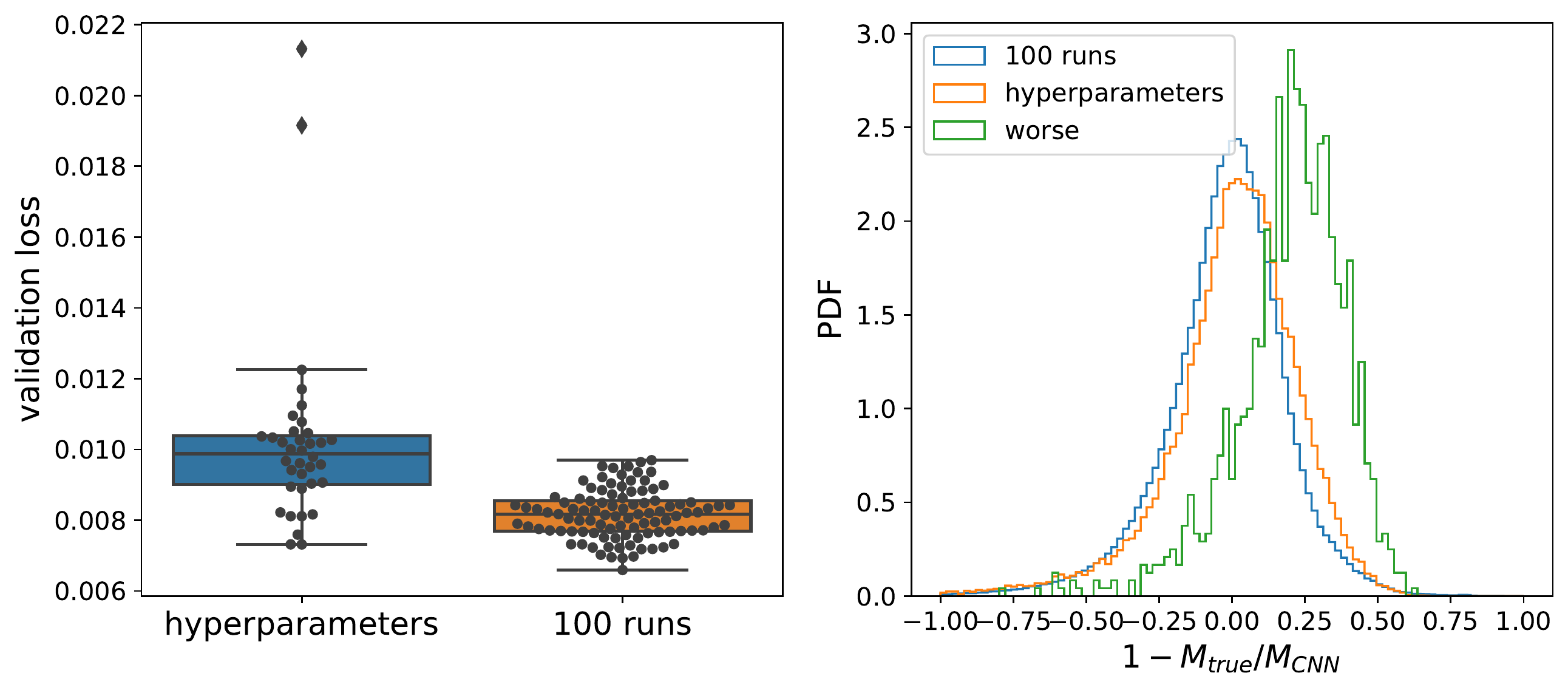}
\caption{Left panel: Validation loss function for the set of hyperparameters and our model with 100 runs. The box (shaded regions) represents the $1-\sigma$ percentiles of the dataset while the whiskers extend to show the rest of the distribution, except for points determined to be “outliers”. Right panel: The distribution of the relative error (x-axis) for the 100 runs (blue), all the models in the hyperparameters space (orange) and the worse set of hyperparameters model (green).}
\label{fig:hypervalidation}
\end{figure}

In  \autoref{fig:hypervalidation}, we show in the left panel the validation loss for the set of hyperparameters and for the 100 runs using our model. Note that the validation loss in the 100 runs is scattered and one cannot conclude that the best run might be better than the best hyperparameter values. In the right panel we show their error distribution defined in Equation (1) of the main article. For the 100 runs the median is -0.01 and the $16^{\text{th}}$ and $84^{\text{th}}$ percentiles are $-0.20$ and $0.16$ respectively. The standard error is $\sigma/\sqrt{N}=0.003$. If we compare these values with respect to the values in \autoref{table:statistics} of the Supplements for $z<0.1$ for \mockdata{} the distribution is similar to the values presented in the Table. However, as shown in \autoref{fig:threshold} some models are highly biased in some mass bins ($b>0.05$). Nevertheless the bias for the average of our 100 models is less than $5\%$. However, almost 50 of these 100 do not meet the validation criteria depicted in \autoref{fig:threshold} for $z<0.1$.  The hyperparameters statistics is similar to the 100 models (orange curve). Nevertheless, for some values of the hyperparameters (green curve) the validation loss is high and therefore these hyperparameter values should not be considered since they fail at capturing the complexity of the data. To properly account for epistemic errors, one has to marginalise over posterior distributions of both parameters and hyperparameters and thus, this is clearly a limitation of the current modelling.

\section{The relative errors} \label{supply:c}
\begin{figure*}
\includegraphics[width=1\textwidth]{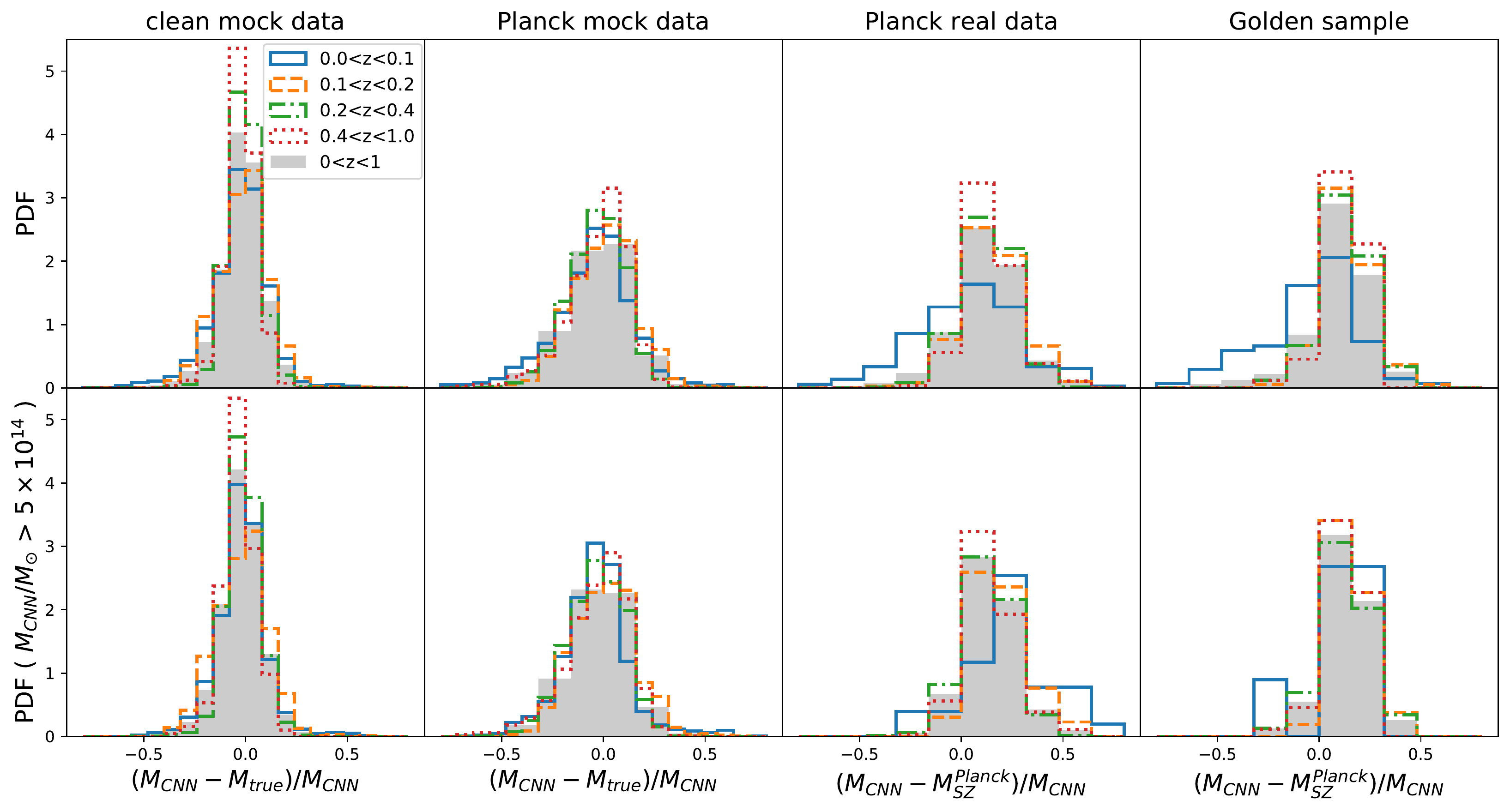}
\caption{Relative error PDFs as defined in Equations (1) and (2) of the main article. The lines represent the relative error for different redshift bins and the gray shaded region corresponds to all the redshifts. In the first row we show the PDFs for all masses while in the second row the data is sampled such that  $\MCNN/\Msun>5\times10^{14}$. The statistics for these PDFs is shown in table \ref{table:statistics}}
\label{fig:planckpdfs}
\end{figure*}

The relative error defined in Equation (1) of the main article can be  interpreted as a probability distribution function (PDF), which contains the intrinsic scatter of our predictions. These PDFs are shown in \autoref{fig:planckpdfs}  for all the data sets described in section method section and for different redshift bins. Furthermore, in the bottom panel we also show the PDFs for massive clusters objects $\MCNN/\Msun>5\times10^{14}$. We also show a table with the values of the PDFs distributions (\autoref{table:statistics}) and also the values of the biases with the standard error computed as $\bP=\text{mean}(\bP)\pm\sigma/\sqrt{N}$, where N is the number of clusters.

 
According to our results, Planck mass estimates $\MPlanck$ and $\MCNN$ are in agreement (up to ~14\% bias with an overall scatter of 30\%). The systematics observed between the two inferred masses can be caused by several factors: 1) the hydrostatic mass bias; 2) our \mockdata{} may be not fully mimicking the \realdata{}; 3) the CNNs are not perfectly performing (an optimal architecture is not found or CNNs in general have limitations to address this problem); 4) the $Y-M$ scaling relation parameters used to compute $\MPlanck$. The first possibility cannot explain the bias mass dependence because in \theth{} simulation the hydrostatic mass bias is almost independent on mass and redshift ranges \citep{giulia300}. The second might be improved by updating the smoothing and noising procedure described in method section. The third problem could be addressed by refining the CNN architecture and training procedure, or maybe a totally different architecture might be an explanation for that. The fourth possibility could explain the bias $\bP$ dependence in mass simply because the Planck scaling relation parameters used in the computation of $\MPlanck$ are not consistent with The300 $Y-M$ scaling law . The difference between the mass estimated using scaling relations and the CNN mass is discuss in section 2 and in the conclusions of the main article. 

 To compare the error with other theoretical works, we define the residual logarithmic mass as \begin{equation}
 \epsilon = \log \left(\MCNN/\Mtrue\right)\text{ .}
\end{equation}
The distributions of this quantity $\epsilon$ can be found in \autoref{fig:zintervals} for different redshift intervals. The distribution corresponding to this work (4 redshifts) has an average standard deviation in the $\epsilon$ distribution of $\sigma = 0.05$ dex which correspond to a $\sim 12\%$ relative error scatter in our clean data set. However, this distribution is not fully Gaussian as can be seen in the skewness and kurtosis values. Nevertheless, our performance in the noisy \mockdata{} has a standard deviation in the $\epsilon$ distribution of $\sigma=0.07$ dex, which corresponds to $\sim 20\%$ scatter in the relative error. Moreover, in \cite{yan2020galaxy} they have used the BAHAMAS simulation to train another VGGNet-based architecture using mock Compton-$y$ parameter maps with value of the standard deviation of also $\sigma = 0.07 $ dex. In \cite{gupta2020massSZ}, they have used an U-Net-based network called (mRestUNET) trained on azimuthally symmetric SZ maps. Their performance on their azimuthally symmetric SZ maps is comparable with the error scatter of $\leq 20\%$. However, their performance on hydrodynamic simulations has a standard deviation of $\sigma\sim 0.23$ dex in the $\epsilon$ distribution and the result is consistent with no mass bias. 

Furthermore, our CNNs performance can also be compared with classical benchmark results corresponding to the $Y-M$ scaling relations. Using the fitted scaling relation of \theth\ simulations shown in Figure 3 of the main article, the scatter of the relative error distribution is $\leq 7\%$. This small scatter is due to the fact that we have integrated the signal to the known aperture $R_{500}$ in simulations. A more realistic approach using SPT \citep{SPTcatalog} tSZ maps with an uncertainty of a single cluster prediction of $\sim 24\%$ is given by \citep{bleem2015galaxy}. This shows that our analysis yields to similar results to these works using $Y-M$ relations. Gravity lensing is expected to provide an unbiased mass. Strong lensing can provide much accurate cluster mass estimates, which is, however, limited by the little number of clusters with arcs and the estimated masses are mostly around the cluster centre \citep[for example][and references therein]{Postman2012}. Weak lensing masses with fewer samples will be dominated by the statistical uncertainties caused by the intrinsic source ellipticity \citep[see][for example]{Hoekstra2013}. Similarly, the richness-based masses have $\sim$0.16 dex errors \citep{Andreon2016,Oguri2014}. Cluster masses estimated with both HE and velocity dispersion methods ($\sim 35\%$ scatter, \cite{Andreon2017}) are biased due to the HE assumption and tracers, respectively. Furthermore, they also tend to be affected by the cluster dynamical state \cite{giulia300}.

Note however  that a more careful modelling in Deep Learning is required in order to realistically model posterior uncertainties. This means that the empirical scatter through cross-validation cannot be misunderstood with a rigorous statistical modelling of uncertainties. Total uncertainties can be understood as the addition of aleatoric and epistemic uncertainties. We have checked  that our model is robust against aleatoric uncertainties by simply varying the number of training samples. However, in order to fully capture epistemic uncertainties one also needs to include the marginalisation over posteriors on the model parameters conditioned on training data. Note that the current analysis does not investigate uncertainties over these parameters and this could impose some  limitations to the model.


A possible way to address this issue is by  implementing some sort of  uncertainty reconstruction methods such as 'Approximate Bayesian Networks' \citep{ho2021approximate} or 'Simulation-based Inference' (SBI) \citep{kodi2021simulation}. We use the later  in the next section.

\section{Uncertainty estimation}
\begin{figure}
\includegraphics[width=\columnwidth]{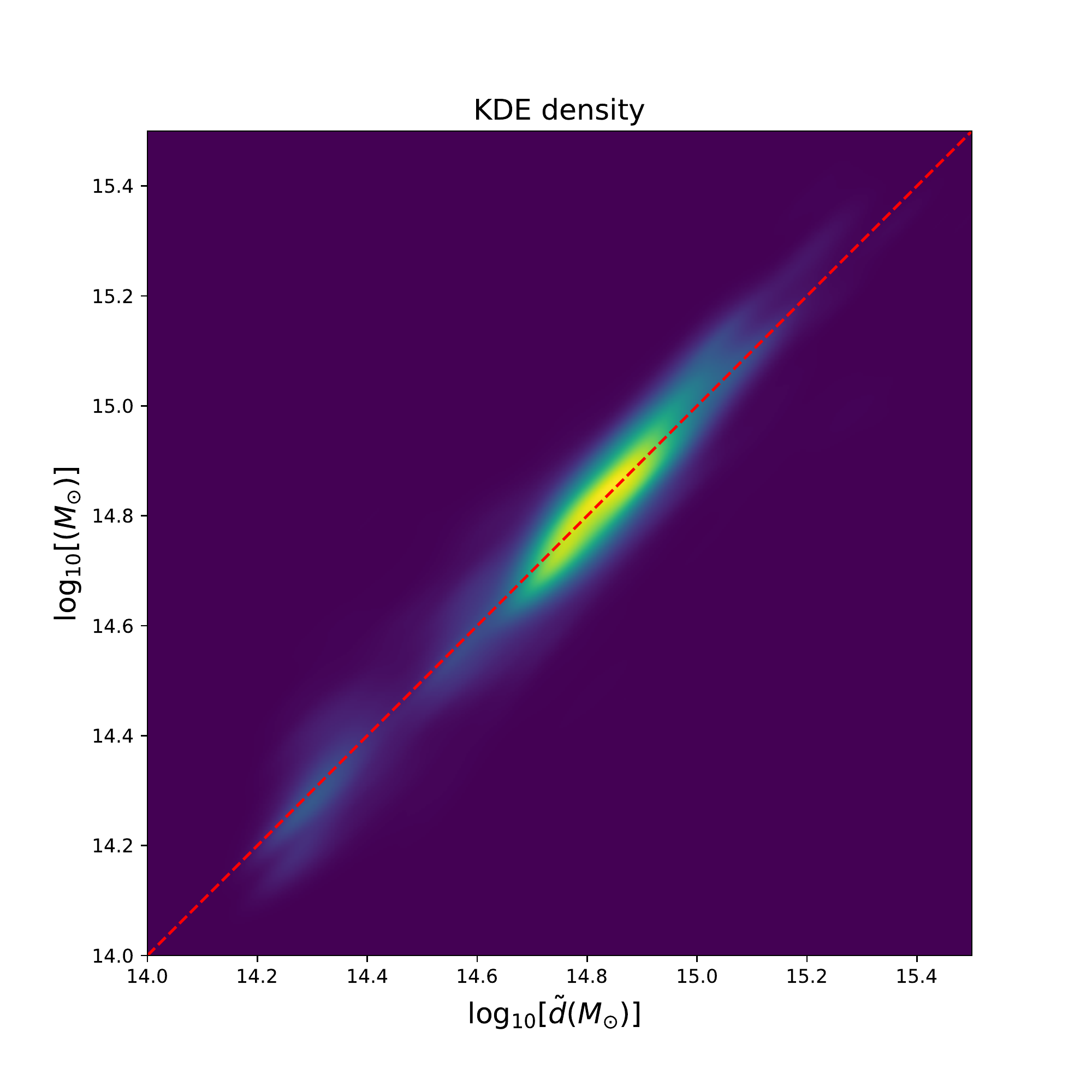}

\caption{KDE density obtained by applying a Gaussian KDE on the dynamical mass estimates $\tilde{d}$ (x-axis) and the true simulated masses $\Mtrue$ (y-axis). We have used a bandwidth scaling factor of 0.2 and the KDE is exclusively applied on the test set.}
\label{fig:kde}
\end{figure}

Our current modelling cannot account for uncertainties, i.e. we only studied how the CNN predictions are compared with the true masses in our simulations, and from that we have studied the relative error statistics defined in Eq. (1). Nevertheless, the current model can be generalised to account for posterior uncertainties in the framework of Simulation-Based Inference  \cite{kodi2021simulation}, which is inspired in the work of \cite{charnock2018SBI}. Given the input data $d$ (SZ mock maps), SBI provides a method to predict the true targets $\tau$ including reliable uncertainties. A CNN has been previously trained to predict the mass by finding the combination of weights $\hat{\theta}$ and hyperparameters $\hat{\gamma}$ that gives the best performance in the sense of minimising the MSE, i.e. 
\begin{equation}
    \text{CNN}(\hat{\theta},\hat{\gamma}): d \rightarrow \tilde{d} \text{ .}
\end{equation}
Here, $\tilde{d}$ is a set of predicted summaries which correspond to the predicted cluster masses. The approximate posterior $\mathcal{P}(M|\tilde{d_{\text{o}}})$ given a set of observed data is then obtained by slicing the joint distribution computed via a Kernel Density Estimator (KDE).  In Figure \autoref{fig:kde}, a Gaussian KDE is used to compute the probability distribution given the  true  masses $M$ and the CNN predictions $\tilde{d}$ for the test set, i.e. $\mathcal{P}(\tilde{d},M)$. Then, a slice $\mathcal{P}(\tilde{d},M)$ at a particular observed CNN prediction $\tilde{d}_{o}$ gives the approximate posterior distribution:
\begin{equation}
    \mathcal{P}(M|d_{o},\tilde{\theta},\tilde{\gamma})\approx \mathcal{P}(M|\tilde{d_{\text{o}}}) \text{ .} 
\end{equation}
Therefore, for any galaxy cluster image of the  Compton-y parameter we have estimated a predicted mass $\MCNN$ with the corresponding uncertainties using the SBI framework. It is important to note that SBI framework is an empirical estimation of the total uncertainties associated to  our mass predictions, guided by the masses in the training set.

The data products, including cluster masses with uncertainties are publicly available  in  our repository at  \url{https://github.com/The300th/DeepPlanck}.

\section{Justification of the selection of redshift bins} \label{sec:redbin}
  Although we have considered dividing our data set only in four redshift intervals to train our CNNs, this division could have been performed differently. The particular chosen number of redshift intervals has to be a trade-off  between the number of snapshots, because the data stored in \theth{} corresponds to a particular value of the redshift (the redshift in our simulation is a discrete variable) and the number of objects inside the interval. To this end, we imposed 1) a minimum of 4 snapshots per redshift interval and 2) a minimum of 5 000 maps, similar to the size of the data set used in \cite{ntampaka2019deep} and \cite{yan2020galaxy}. The first condition implies that the minimum size of the redshift interval has to be 0.1, provided the fact that there are 4 snapshots every 0.1 redshift increase. Together with the second condition, the number of possible intervals is 8. These intervals are : $0< z\leq 0.1$,  $0.1<z\leq 0.2$,  $0.2< z\leq 0.3$,  $0.3< z\leq 0.4$,  $0.4< z\leq 0.5$,  $0.5< z\leq 0.6$,  $0.6< z\leq 0.7$ and  $0.7< z\leq 1$.
  
  Furthermore, we have repeated the same validation procedure training as many CNNs as the following redshift intervals:
  \begin{itemize}
      \item 1 interval) $z\in(0,1]$
      \item 2 intervals) $z\in (0,0.2], (0.2,1]$
      \item 4 intervals) $z\in (0,0.1],(0.1,0.2],(0.2,0.4],(0.4,1]$
      \item 8 intervals) $z\in (0,0.1],(0.1,0.2],(0.2,0.3],(0.3,0.4],(0.4,0.5],(0.5,0.7],(0.6,0.7],(0.7,1.0]$
  \end{itemize}
  
\begin{figure}
\center
\includegraphics[width=0.7\columnwidth]{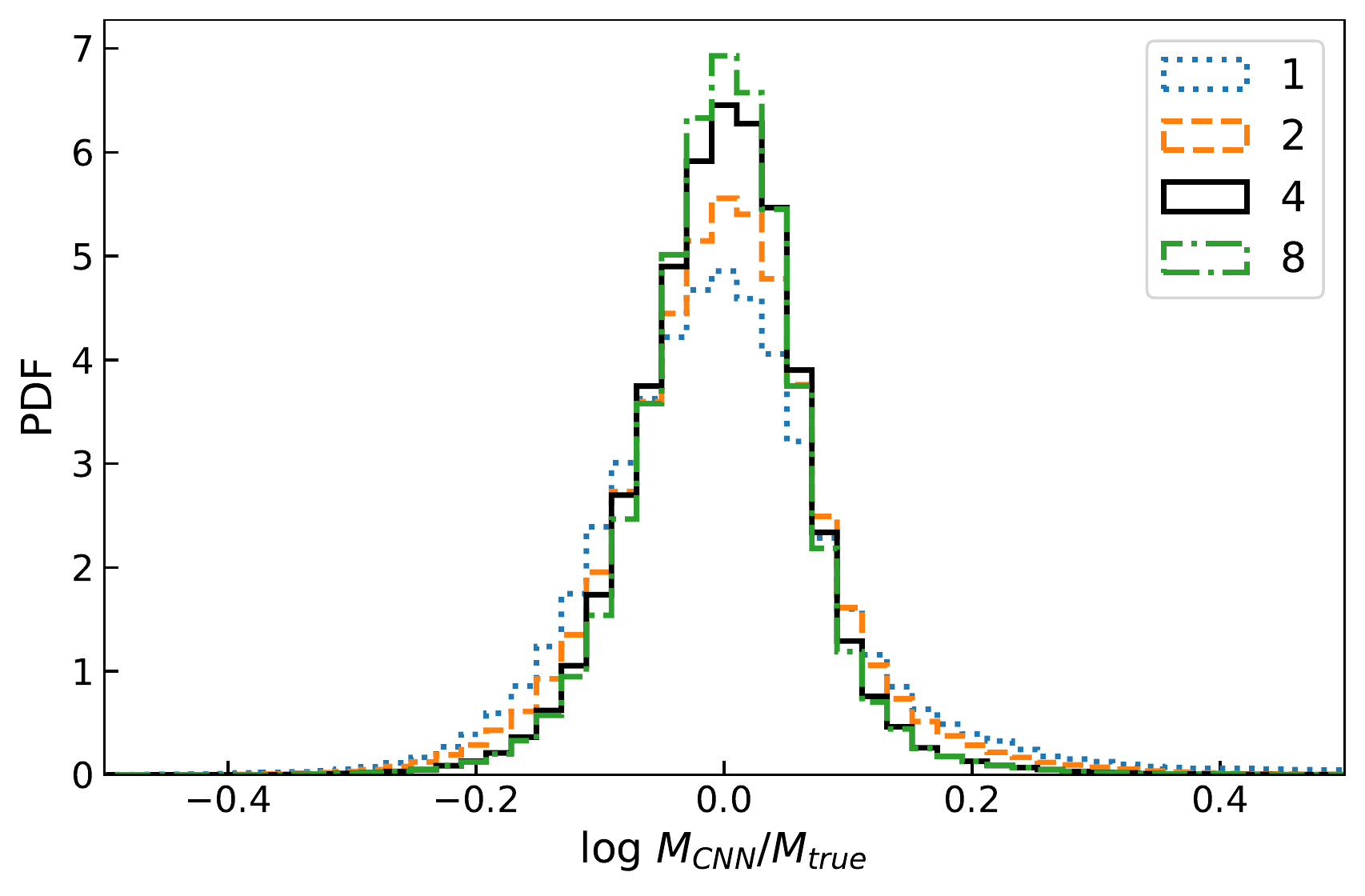}

\caption{PDF corresponding to the logarithmic difference of the predicted mass $\MCNN$ and the true mass $\Mtrue$ for the different considered redshift intervals. The 4 moments of the distributions are given in \autoref{table:zintervals}.}
\label{fig:zintervals}
\end{figure}

In \autoref{fig:zintervals}, we show the probability distribution function (PDF) of the logarithmic residuals $\log \MCNN/\Mtrue$ for the different considered redshift ranges. The mean, standard deviation (std), skewness and kurtosis for these PDFs are listed in \autoref{table:zintervals}. As a general result, the scatter of $\log \MCNN/\Mtrue$ decreases as the number of redshift intervals increases. We also notice that considering 4 intervals instead of 8 is not relevant. It is also interesting to note that the skewness is positive for all the models, which indicate that the tail of the distribution is always on the right. Furthermore, the kurtosis increases with the number of redshift intervals because the standard deviation decreases and the outliers are further away from the width of the distribution. Therefore, these distributions have more outliers than a normal distribution. We have to point out that this analysis has been performed using the \mockdata{}{}.

\begin{table*}
\centering
\caption{Mean, standard deviation (std), skewness(0) and kurtosis(3) for the distributions shown in \autoref{fig:zintervals}. The values shown inside the brackets represent the standardised moments for a Gaussian distribution. }
\begin{tabular}{ccccc} 
\hline
redshift intervals & mean & std    & skewness(0) & kurtosis(3)  \\ 
\hline
1                  & -0.0002  & 0.1078 & 0.8842      & 6.7942       \\
2                  & -0.0002  & 0.0879 & 0.3535      & 5.6138       \\
4                  & -0.0014  & 0.0710 & 0.3978      & 7.2888       \\
8                  & -0.0009  & 0.0693 & 0.4301      & 7.8559       \\
\hline
\end{tabular}
\label{table:zintervals}
\end{table*}

\setlength{\tabcolsep}{1pt}
\renewcommand{\arraystretch}{1.5}

\begin{table*}
\centering
\caption{We show the median and the $16^{\text{th}}-84^{\text{th}}$ percentile of the PDFs in \autoref{fig:planckpdfs} for different redshift ranges and data sets. The results are shown in the following format: $\text{median}^{+|84^{\text{th}}-\text{median}|}_{-|16^{\text{th}}-\text{median}|}$ $(\pm \sigma/\sqrt{N})$. Note that in brackets we provide the standard error. The number of maps N of \realdata{} and \goldendata{} is also shown. The number of clusters for \cleandata{} and \mockdata{} are  roughly 74. The data is averaged considering all clusters corresponding to the redshift and mass indicated.}
\label{table:statistics}
\footnotesize
\begin{tabular}{ccccccc} 
\hline
Redshift & \cleandata{} & \mockdata{} & \realdata{} & N & \goldendata{} & N \\
\hline
(0, 0.1] & $-0.03^{+0.10}_{-0.13}$ ($\pm0.002$) & $-0.05^{+0.15}_{-0.20}$ ($\pm0.003$)  & $0.03^{+0.23}_{-0.27}$ ($\pm0.018$) & 228  & $-0.06^{+0.15}_{-0.31}$ ($\pm0.029$)  & 87   \\
(0.1, 0.2] & $-0.01^{+0.10}_{-0.13}$ ($\pm0.002$) & $0.01^{+0.13}_{-0.17}$ ($\pm0.002$) & $0.15^{+0.13}_{0.12}$  ($\pm0.008$) & 245  & $0.13^{+0.15}_{-0.0.8}$ ($\pm0.011$) & 103  \\
(0.2, 0.4] & $-0.01^{+0.07}_{-0.08}$ ($\pm0.001$)  & $-0.04^{+0.13}_{-0.15}$ ($\pm0.002$) & $0.13^{+0.13}_{-0.11}$ ($\pm0.006$)& 443  & $0.13^{+0.12}_{-0.11}$ ($\pm0.010$) & 150  \\
(0.4, 1] & $-0.03^{+0.06}_{-0.07}$ ($\pm0.001$) & $-0.03^{+0.11}_{-0.16}$ ($\pm0.002$)  & $0.12^{+0.09}_{-0.10}$ ($\pm0.009$) & 178  & $0.12^{+0.07}_{-0.07}$ ($\pm0.012$)  & 55   \\
(0, 1] & $-0.02^{+0.09}_{-0.10}$ ($\pm0.001$) & $-0.03^{+0.14}_{-0.17}$ ($\pm0.002$) & $0.11^{+0.14}_{-0.15}$ ($\pm0.005$) & 1094 & $0.08^{+0.11}_{-0.13}$ ($\pm0.009$)   & 395  \\ 
\hline
\multicolumn{7}{c}{$ \MCNN >5\times 10^{14}\ M_{\odot}$}\\ 
\hline
(0, 0.1]  & $-0.02^{+0.09}_{-0.10}$ ($\pm0.002$) & $-0.04^{+0.12}_{-0.14}$ ($\pm0.003$)  & $0.26^{+0.30}_{0.18}$ ($\pm0.043$)  & 33   & $0.10^{+0.07}_{-0.14}$ ($\pm0.057$)  & 7    \\
(0.1, 0.2]   & $-0.02^{+0.11}_{-0.14}$ ($\pm0.002$) & $-0.01^{+0.14}_{-0.16}$ ($\pm0.002$)  & $0.16^{+0.14}_{-0.11}$ ($\pm0.014$) & 82   & $0.15^{+0.10}_{-0.10}$ ($\pm0.017$) & 33   \\
(0.2, 0.4] & $-0.01^{+0.08}_{-0.07}$ ($\pm0.001$)   & $-0.04^{+0.13}_{-0.15}$ ($\pm0.002$) & $0.13^{+0.12}_{-0.11}$ ($\pm0.014$)& 402  & $0.13^{+0.11}_{-0.12}$ ($\pm0.009$)  & 145  \\
(0.4, 1] & $-0.03^{+0.07}_{-0.07}$ ($\pm0.002$) & $-0.03^{+0.12}_{-0.16}$ ($\pm0.003$)  & $0.14^{+0.09}_{-0.01}$ ($\pm0.011$) & 178  & $0.12^{+0.07}_{0.07}$ ($\pm0.012$) & 55 \\
(0, 1] & $-0.02^{+0.09}_{-0.09}$ ($\pm0.001$) & $-0.03^{+0.14}_{-0.15}$ ($\pm0.002$)  & $0.14^{+0.12}_{-0.11}$ ($\pm0.008$)   & 695  & $0.13^{+0.09}_{-0.11}$ ($\pm0.010$) & 240  \\ 
\hline
\multicolumn{7}{c}{$\MCNN <3\times 10^{14}\ M_{\odot}$}   \\ 
\hline
(0, 0.1] & $-0.03^{+0.10}_{-0.15}$ ($\pm0.003$) & $-0.05^{+0.16}_{-0.24}$ ($\pm0.005$)  & $-0.03^{+0.24}_{-0.27}$ ($\pm0.021$) & 164  & $-0.08^{+0.16}_{-0.31}$ ($\pm0.021$) ($\pm0.033$)& 74   \\
\hline
\end{tabular}
\end{table*}

\section{Interpretability}
Machine learning models are usually referred to as Black box estimators and it is not trivial to understand how they make their predictions. To address this problem, we compute the gradient of the output neuron (the predicted mass $\MCNN$) with respect to image input pixels. This shows which pixels are activating for a given image when it comes to inferring the mass of the clusters. This interpretation algorithm is a generalised version of  Google's {\it  Deep Dream} algorithm \citep{DeepDream} for regression and it has been previously studied by \citep{ntampaka2019deep} for X-ray mock maps  and \citep{yan2020galaxy} in the case of X-ray, SZ and optical maps. In order to be able to visualise which part of the images the CNN is focusing on, we defined a `saliency map' as the absolute value of the gradient whose pixel values are shifted between 0 and 1: 0 for not activated and 1 for the maximum activation. 

\begin{figure}
\includegraphics[scale=0.6]{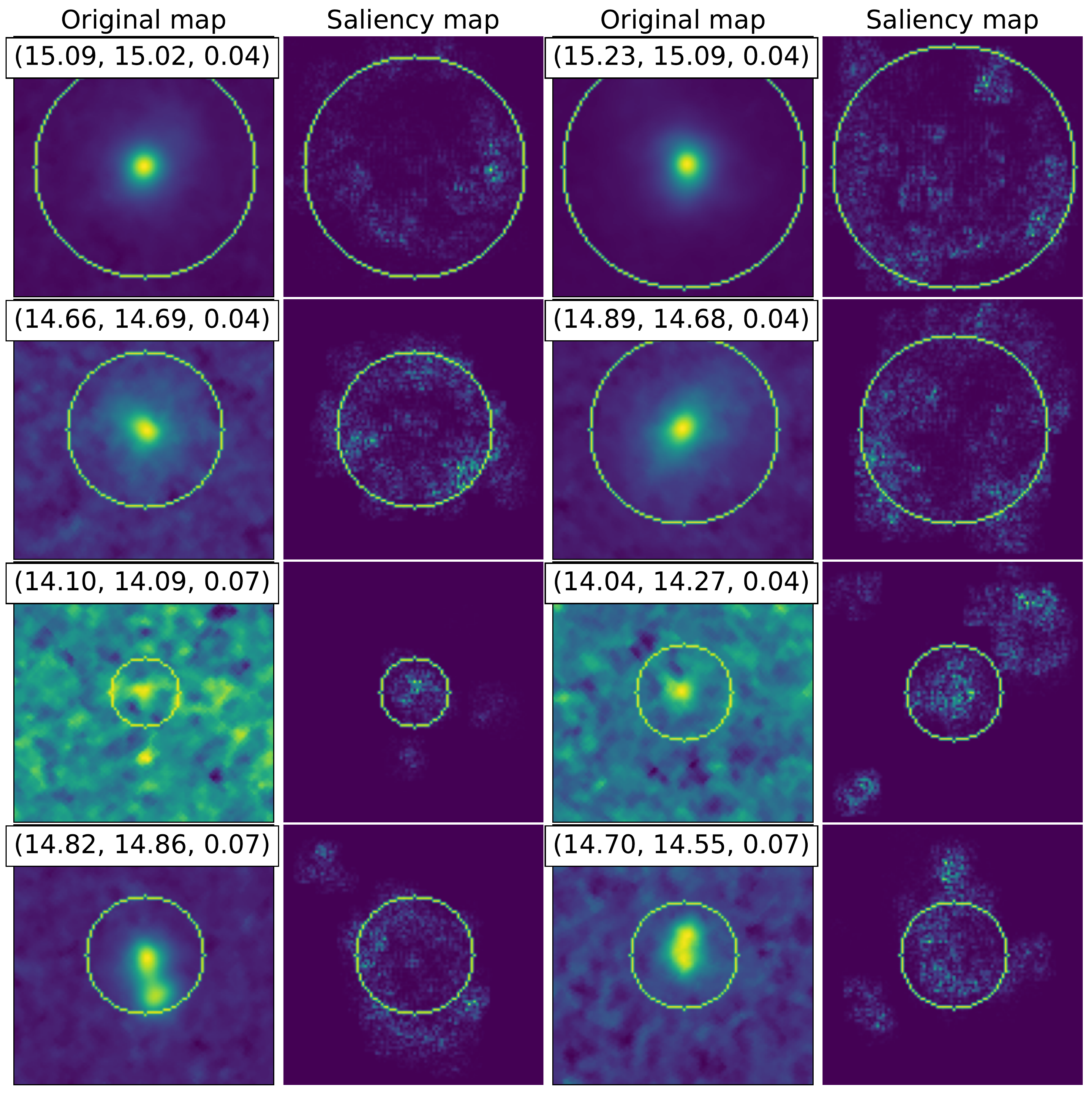}
\caption{Left column: Paired original-saliency maps corresponding to the \mockdata{} for well predicted masses. Right column:  Same as left column for inaccurately predicted masses. From top to bottom we show maps corresponding to clusters with different masses labelled as $(\log \Mtrue,\log \MCNN,\text{redshift})$.  The bottom panel represents an example of a merger event. IN all maps, green circles correspond to $R_{500}$.}
\label{fig:heatmaps}
\end{figure}

We show eight examples of `saliency maps' in \autoref{fig:heatmaps}. The two columns on the left correspond to paired original-saliency maps of \mockdata{} for clusters with well-predicted masses while the two columns on the right for clusters with inaccurately predicted masses. We include true masses, predicted masses by CNN and redshifts inside these boxes $(\log \Mtrue, \log\MCNN, \text{redshift})$. From top to bottom, different examples varying the mass of the clusters are displayed. The last example at the bottom depicts a merger event. Note that the gradients are activated following the shape of the merger event, i.e. the gradients do not have a perfect circular shape. 
To our understanding, the inaccurately predicted masses can be caused by different reasons: the improper `saliency maps', for example the low mass cluster and merger cluster on the third and fourth rows; the limitation of the CNN due to less samples at the most massive cluster mass, the first and the second row; or some other properties of these maps, especially the signal at around $R_{500}$.

\begin{figure}
\center
\includegraphics[width=0.8\columnwidth]{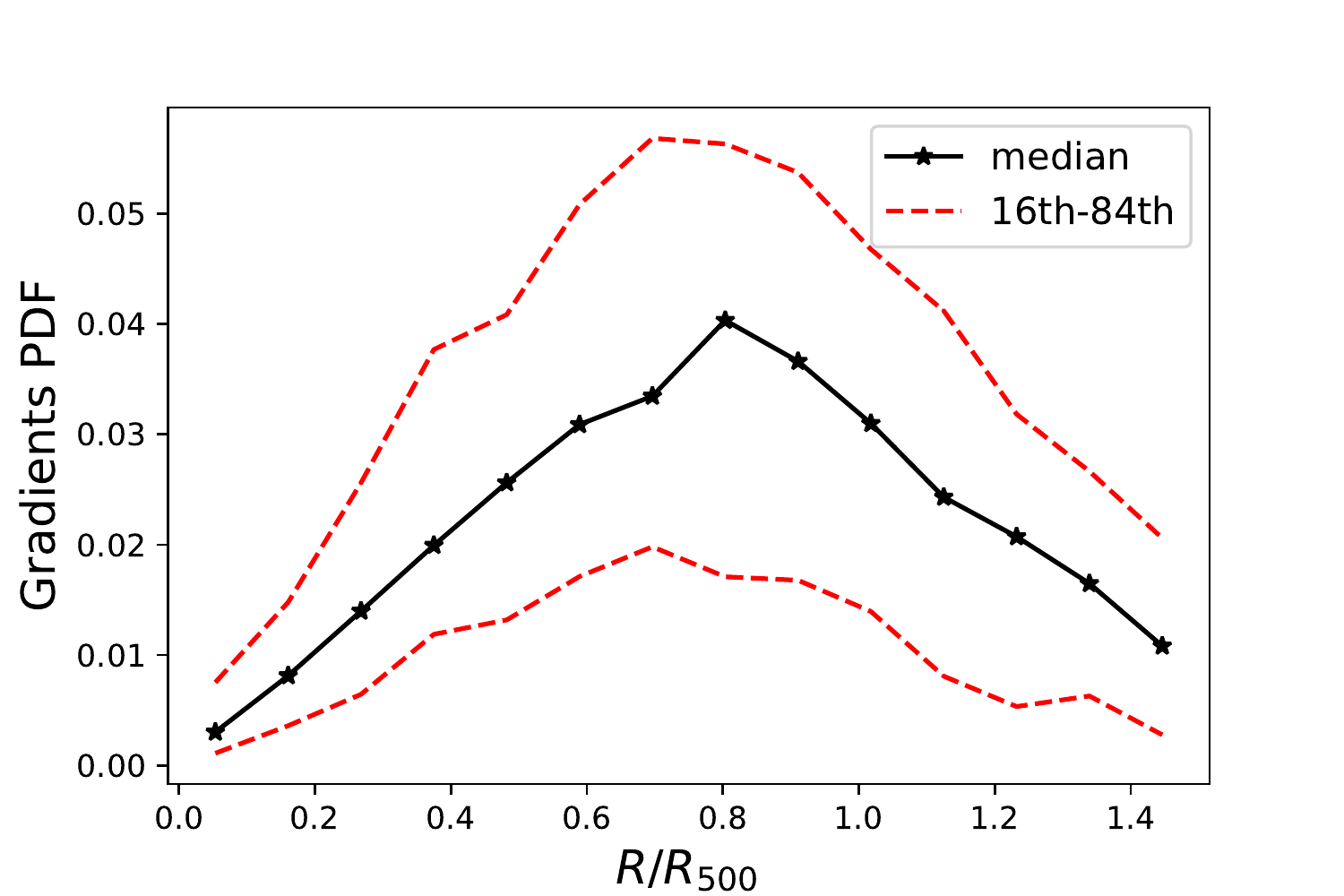}
\caption{Gradient PDF distribution as a function of radius, normalised to $R_{500}$. We show the median values (black points) and $16^{th}$  and $84^{th}$  percentiles (dashed red lines)}
\label{fig:gradientPDF}
\end{figure}

In order to analyse more quantitatively the information of the saliency maps, we compute their radial distribution as a function of $R_{500}$. In \autoref{fig:gradientPDF}, we show the median, the $16^{th}$ and $84^{th}$ percentiles of  all the stacked saliency maps. For this purpose all saliency maps are normalised such that the sum of all the pixels inside each map is one. As a general result, CNN predictions mostly ignore the noise  and the most important contribution comes from pixels between $0.4-1.2 R_{500}$. 

 By applying this algorithm, we found no significant difference when computing the gradients in \realdata{} and \goldendata{}. This technique is based on a qualitative interpretation of CNN results based on feature attribution. A  more quantitative interpretation  can be done using the {\it Testing with Concept Activation Vector} \citep[TCAV][]{Kim2017}, but  it  is left for future work.
 


\begin{figure}
\center
\includegraphics[width=0.8\columnwidth]{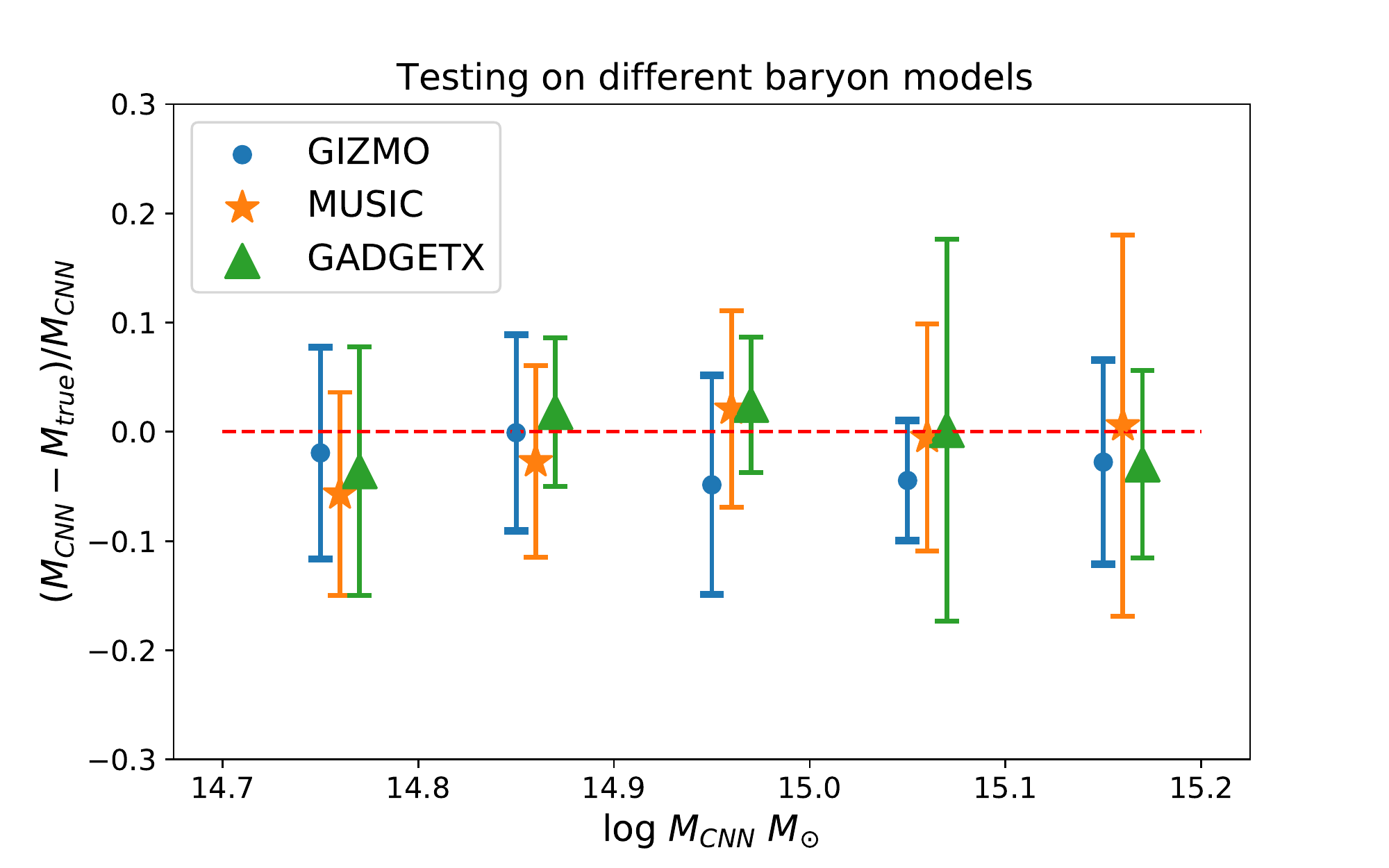}
\caption{ Relative difference $1-\Mtrue/\MCNN$ (y-axis) as a function of $\MCNN$ (x-axis) for our model trained on {\gadgetx} testing on  data from simulations  with different baryonic physics. Points correspond to the mean values and error bars represent the standard deviation for the different mass bins.}\label{fig:baryonmodels}
\end{figure}

In this letter, We have claimed that the CNN trained with one baryon model (\gadgetx) can be applied to mock maps from the other models (\gadgetmusic{} or \simba). After examining these activation maps, we suggest that the reasons for that are (1) the global properties, such as the integrated $Y$ and $M_{500}$, thus the $Y-M$ relation, are very similar between these models \citep{Cui2018,Cui2022}; (2) as shown in \autoref{fig:heatmaps} in this supplement, the pixel contributions are mostly coming  from $0.4 - 1.2 \times R_{500}$. The pixels at the core of the clusters, though having higher S/N and larger difference between these hydrodynamic models, actually contribute little to the   cluster mass estimation. Therefore, it is not surprising to see that the CNN is not significantly biased towards  different baryon models. In order to perform this test, we have trained  the same CNN model with clusters from \theth\ \gadgetx\  simulations at redshift z=0 but at  high resolution (5 arcsec). The results are presented in \autoref{fig:baryonmodels} corresponding to cluster mass predictions from   the {\simba}, {\gadgetmusic} and {\gadgetx} maps. Note that the model has only been trained with {\gadgetx} data.  At the time of writing this paper, we only had available data from {\simba} for the most massive (central) halos and the test is valid for $\log M/\Msun \gsim 14.7$ or equivalently $M\gsim 5\times 10^{14} \Msun $. In \autoref{table:baryonmodels} we show the value of the mean bias for our 3 simulations together with their standard error. As a general result, {\simba} and {\gadgetx{}} have the same bias (totally different baryon models although both have AGN feedback implemented) but the bias in {\gadgetmusic} is lower by $\sim 0.03\%$ (different subgrid physics and no AGN feedback). These results suggest that baryon physics has little effect on the integrated Compton-y parameter.

\begin{table}
\centering
\caption{Average bias with standard error for the inferred masses from  maps of  three different simulations. The CNN has been trained on data exclusively from  The300 {\gadgetx{}} simulation}
\label{table:baryonmodels}
\begin{tabular}{ccc} 
\toprule
simulation & mean bias & standard error  \\ 
\hline
{\gadgetmusic{}}      & -0.068    & $\pm 0.002$           \\
{\simba}      & -0.025   & $\pm 0.002$           \\
{\gadgetx}   & -0.031    & $\pm 0.003$           \\
\bottomrule
\end{tabular}
\end{table}

\end{document}